\documentstyle[12pt,lnfprep,epsfig]{article}
\def\epem{ e^+e^- }
\def\c2{CLEO~II.V}
\def\ccb{ c\bar{c} }
\def\d0d0{ D^0\bar{D}^0 }
\def\p0p0{ P^0\bar{P}^0 }
\def\qp2{ \Bigl| \frac{q}{p} \Bigr|^2 }
\def\pq2{ \Bigl| \frac{p}{q} \Bigr|^2 }
\def\rarr{ \rightarrow }

\def\jp{ J/\psi }
\def\pspr{ \psi^\prime }
\def\Journal#1#2#3#4{{#1} {\bf #2}, #3 (#4)}

\def\PRL{\em Phys. Rev. Lett.}

\def\PRp{{\em Phys. Rept.}  }
\def\PR{{\em Phys. Rev.}  }
\def\PL{{\em Phys. Lett.}  }
\def\RMP{{\em Rev. Mod. Phys.}  }
\def\ARNPS{{\em Ann. Rev. Nucl. Part. Sci.}   }
\def\PPNP{{\em  Prog. in Part. Nucl. Phys.}   }
\def\be{\begin{equation}}
\def\ee{\end{equation}}
\def\bea{\begin{eqnarray}}
\def\eea{\end{eqnarray}}
\newcommand{\Header}{
  \begin{tabular}{rl}
  \hspace{-.4cm}\includegraphics{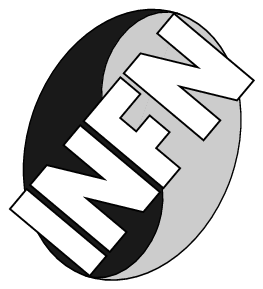} &
    \renewcommand{\arraystretch}{0.5}
    \begin{tabular}{r}
      {\hspace{1cm}~\LARGE\sffamily LABORATORI~ NAZIONALI~ DI~ FRASCATI}\\
      \\
      {\Large\sffamily SIS-Pubblicazioni}\\
    \end{tabular}
    \renewcommand{\arraystretch}{1}
  \end{tabular}
  \vskip 1cm
  \begin{flushright}
  \renewcommand{\arraystretch}{0.5}
    \begin{tabular}{r}
      {\underline{LNF-99/029 (P)}}\\    % insert here the preprint number
      {\small 25 Ottobre 1999} \\      % insert here the preprint Date
      \\
    \end{tabular}
  \end{flushright}
  \renewcommand{\arraystretch}{1}
  \vskip 1 cm
  }
\begin{document}
\begin{titlepage}
\title{
  \Header
  {\large \bf CHARM OVERVIEW}
}
\author{ Stefano Bianco \\
 Laboratori Nazionali di Frascati  \\
via E.~Fermi 40, Frascati 00044, Italy      \\
E-mail: bianco@lnf.infn.it
}
\maketitle
\baselineskip=14pt

\begin{abstract}
This paper is aimed at giving a complete review of the latest (post-1998
conference) experimental results on charm physics.
\end{abstract}

\vspace*{\stretch{2}}
\begin{flushleft}
% insert here the PACS number
  \vskip 2cm
{PACS:13.20.Fc;13.25.Ft;13.30.Eg;13.30.Ce;13.30.-a;13.60.-r;13.60.Rj; \\
13.60.Le;14.20.Lq;14.40.Lb;14.65.DW}  
\end{flushleft}
\begin{center}
Invited review presented at the XIX Physics in Collision, Ann Arbor (USA),
June 1999.
\end{center}

\end{titlepage}
\pagestyle{plain}
\setcounter{page}2
\baselineskip=17pt
 \section{Introduction}
 The study of the  c--quark generates about 200 papers 
 and 800 measurements per year, for a total of 15
 collaborations and  700  physicists  involved.
% Fig.\ref{charm}. 
 Such an  effort has led us to a conflicting 
 situation. On one  hand, the advent of high-statistics, high-resolution
 experiments has turned  c--quark physics into precision physics. On the other
 hand, the c--quark mass scale 
 is too large for chiral symmetry methods, while theories based on
 expansions of heavy  quark masses, such as Heavy Quark Effective
 Theory\cite{Isgur:1989vq} (HQET)
 or Operator Product Expansion\cite{Wilson:1969ey,bi96} (OPE),  are often
 questionable because  $m_c$ may not be large enough. 
 Nonetheless, remarkable agreement is often found when
 such theories make predictions, most notably on semileptonic decays,
 lifetimes, and spectroscopy. 
\par
 The 1998-1999 scenario is full of results -- from new experiments (SELEX
 and BES), experiments that have  
 undergone significant upgrades (FOCUS, \c2, E835), and others that are
 planning 
 upgrades for charm physics (HERMES), experiments at 
 their peak publication rate (E791), and experiments (at LEP and HERA)
 that keep their charm working groups alive and vital.
 Interesting news comes from the neutrino (CCFR, CHARM~II) and heavy-ion
 (NA50) groups. 
% The liveliness of the field is also testified by the number of dedicated
% meetings (cite HQ98 KAON MORIOND HF etc), alongside the traditional
% global conferences such as EPS, PIC, and ICHEP.
\par
 In this paper I  have tried to present a review of today's
 scenario rather than pursue the goal of  detailed investigation.
 In setting the physics scenario for each item, I was  guided
 by recent reviews\cite{qu98,go97,bt96,Cumalat:1996za,wi98}, 
 from which I  borrowed copiously.
 With such a goal in mind, I  apologize in advance to those whose
 work has been   left unmentioned.  
% staqndard comments (footnote see appel's humorous anedoctal...
%
\section{Production mechanisms}
 The QCD picture of charm production consists of  parton-level hard
 scattering, which produces the $\ccb$ pair,  hadronization of the $\ccb$
 into a charmed hadron, and a final stage in which the charmed hadron
 travels through and interacts with the hadronic matter, followed by decay.
 Thanks to the generally large momentum transfers involved, the hard
 scattering is traditionally a   good testing ground for perturbative QCD
 techniques,  while the hadronization stage has the complication of matching
 experimental kinematical  distributions of charm hadrons with predictions
 of a suitable dressing  mechanism. Theoretical updates on both topics were
 given at HQ98 \cite{Olness:1998si,Norrbin:1998jz}.
  In general, any asymmetry is due to hadronization since $\ccb$
 asymmetries  in NLO QCD are very small. For a seminal
 review see ref.\cite{Appel:1992em,Appel:1993hn}.
 \par
\begin{figure}[t]
 \begin{center}
  \vspace{10cm}
  \includegraphics{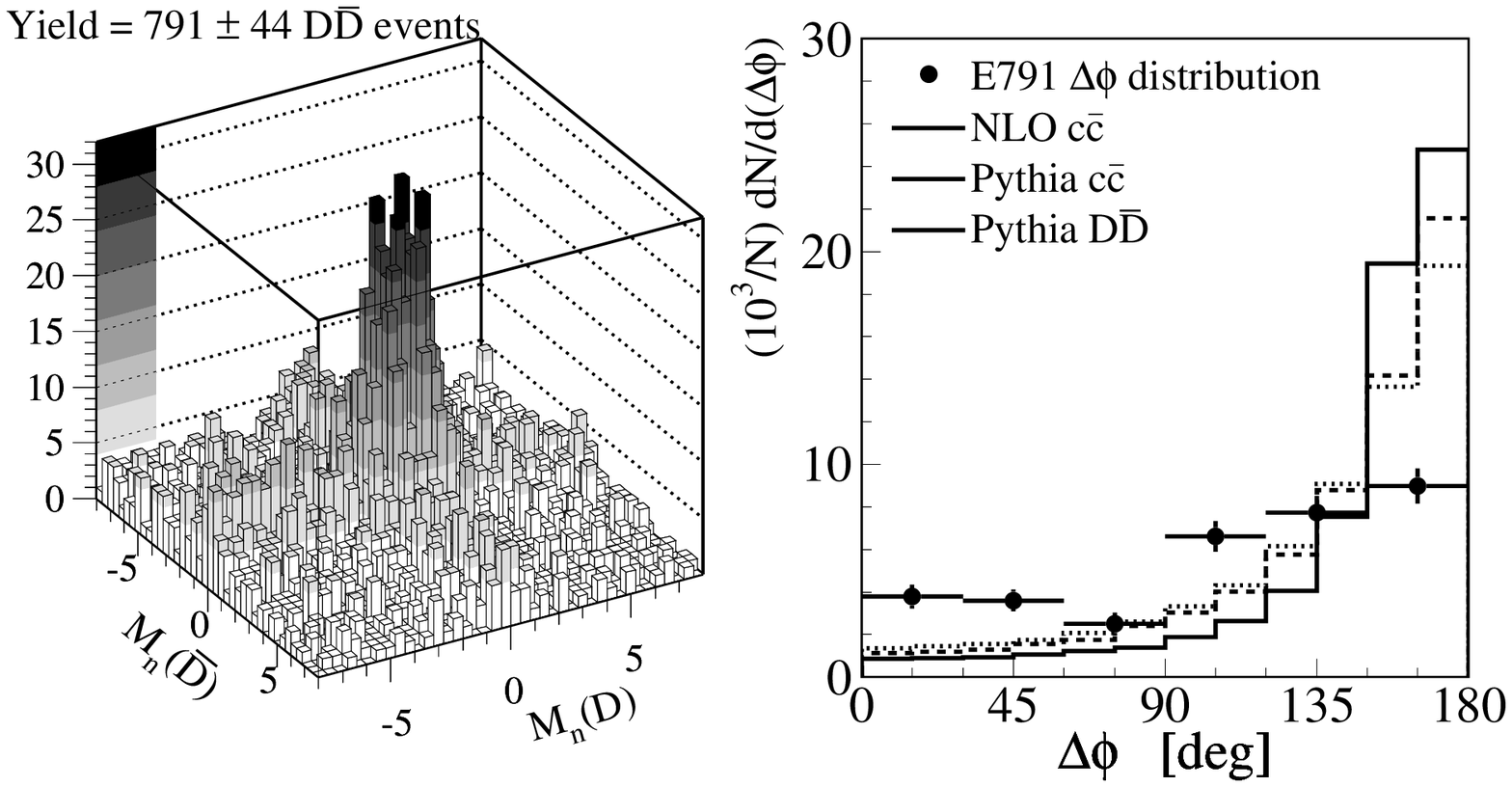}
  \includegraphics{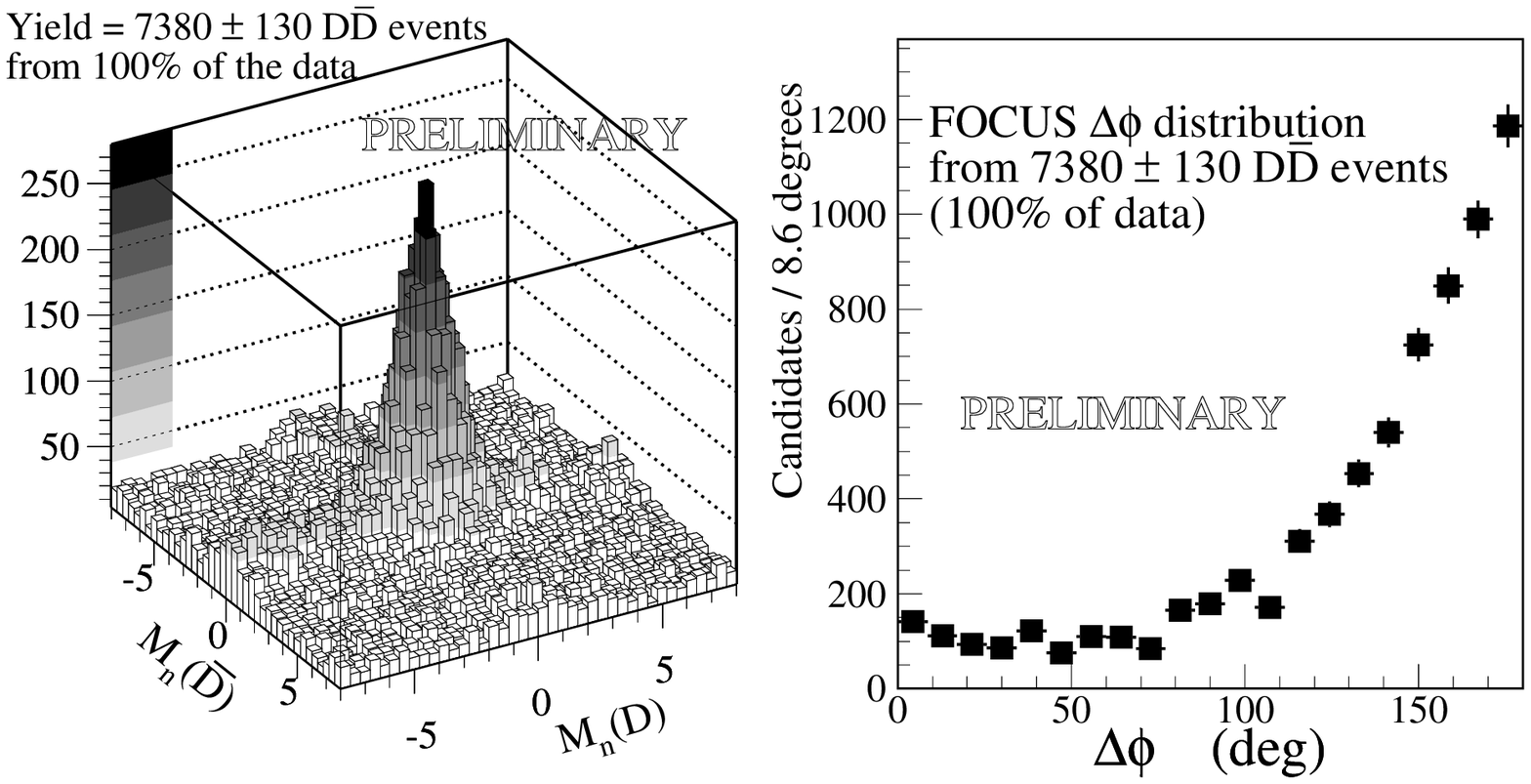}
 \end{center}
 \caption{
  Mass correlation plots (left), and  acoplanarity angle distributions
  (right) for hadro- (E791)  and   photo-produced (FOCUS) $D\bar D$ pairs.
   \label{fig:prodo}
 }
 \vfill
\end{figure}

 A wealth of new results comes from 
 Fermilab fixed-target experiments E791, SELEX and FOCUS (pion, hyperon, and
 photon beams  respectively)\cite{Stenson:1999moriond}. 
 Correlation studies of   fully reconstructed hadroproduced $D\bar{D}$
 pairs by E791\cite{Aitala:1998kh}   show less correlation than that
 predicted by the Pythia/Jetset Monte Carlo event generator for  $D\bar{D}$
 production, which, however, follows the experimental trend better than a
 pure NLO QCD parton level prediction (Fig.\ref{fig:prodo}).
%?????????? CHIEDI A WISS DI FARE OVERLAY
 Coherently,  $D\bar{D}$ correlation in photoproduction is much more pronounced
 because of a lack of hadronization contributions of the projectile remnants.
\par
 \begin{table}
 \caption{Compilation of $\Lambda_c/\bar{\Lambda}_c$
  photoproduction asymmetry measurements 
 (adapted from  ref.$^{13}$).
%\cite{Stenson99}CAVE ???? A MANO STENSON APS). 
   \label{tab:asy}
 }
 \footnotesize
 \begin{center}
\begin{tabular}{|l|r|} \hline
 Experiment  & Asymmetry \\
\hline
 FOCUS prel.             &  $0.14\pm 0.02$ \\
 E687  (93)            &  $0.04\pm 0.08$ \\
 E691 (88)             &  $0.11\pm 0.09$ \\ 
 NA1  (87)              &  $0.14\pm 0.12$ \\ 
\hline
\end{tabular}
  \vfill
 \end{center}
\end{table}
\par
 New results from E791\cite{Stenson:1998mr} also report  
 particle-antiparticle production  asymmetries for $D^{\pm}$, $D^{\pm}_s$
 and $\Lambda_c$, and a cross-section measurement\cite{Aitala:1999yp} with
 a 500-GeV $\pi^-$ beam 
 on a nuclear target of $\sigma(D^0+\bar{D}^0 ; x_F > 0) = 15.4^{+1.8}_{-2.3}
 \mu$b/nucleon. Asymmetry results show clear evidence for leading
 effects. Both measurements are in principle sensitive to $m_c$; in
 practice it is necessary to first nail down a few other parameters
 (factorization scale, intrinsic parton momentum $k_t$, etc). In general,
 E791 asymmetry results favor $m_c=1.7\,{\rm  GeV}/c^2$, while the
 cross-section measurement is compatible with 
 $m_c=1.5\,{\rm GeV}/c^2$. 
 For the renormalization scale and scheme adopted for the above definitions
 of $m_c$ the interested reader is referred to
 reff.\cite{Stenson:1998mr,Aitala:1999yp}.
 SELEX preliminary results\cite{Gottschalk99},
 which  include   comparison  of $\Lambda_c$ and $D^{\pm}$ asymmetries with
 proton, pion, and  hyperon  beams, also indicate large leading
 effects. FOCUS  presented clear evidence\cite{Stenson:1999moriond}  of
 $\Lambda_c$ asymmetry in photoproduction 
 (Tab.\ref{tab:asy}). With a photon beam any asymmetry has to  be attributed
 to the  target, and the picture is consistent with  $\Lambda_c(cud)$
 production being more  probable than $\bar{\Lambda}_c$. When comparing
 results from different photoproduction experiments, it should be also
 kept in mind how  the asymmetry is in general
 $\sqrt{s}$-dependent, as well as dependent on the exact acceptance
 kinematics, i.e., the $x_F$ range, etc.%
\par
 Neutrino charm production also gives estimates of
 $m_c$. Results\cite{Adams1998}  of
 CCFR (Fermilab) and CHARM~II (CERN) experiments provide
 analysis-dependent values: CCFR (next-to-leading order) and CHARM~II find
 $m_c\sim 1.7\,{\rm  GeV}/c^2$, while CCFR (leading order) favors $m_c\sim
 1.3\,{\rm  GeV}/c^2$. New data are expected very soon from the successor 
 CCFR  experiment, NuTeV.
\par
\begin{figure}
 \begin{center}
  \epsfig{file=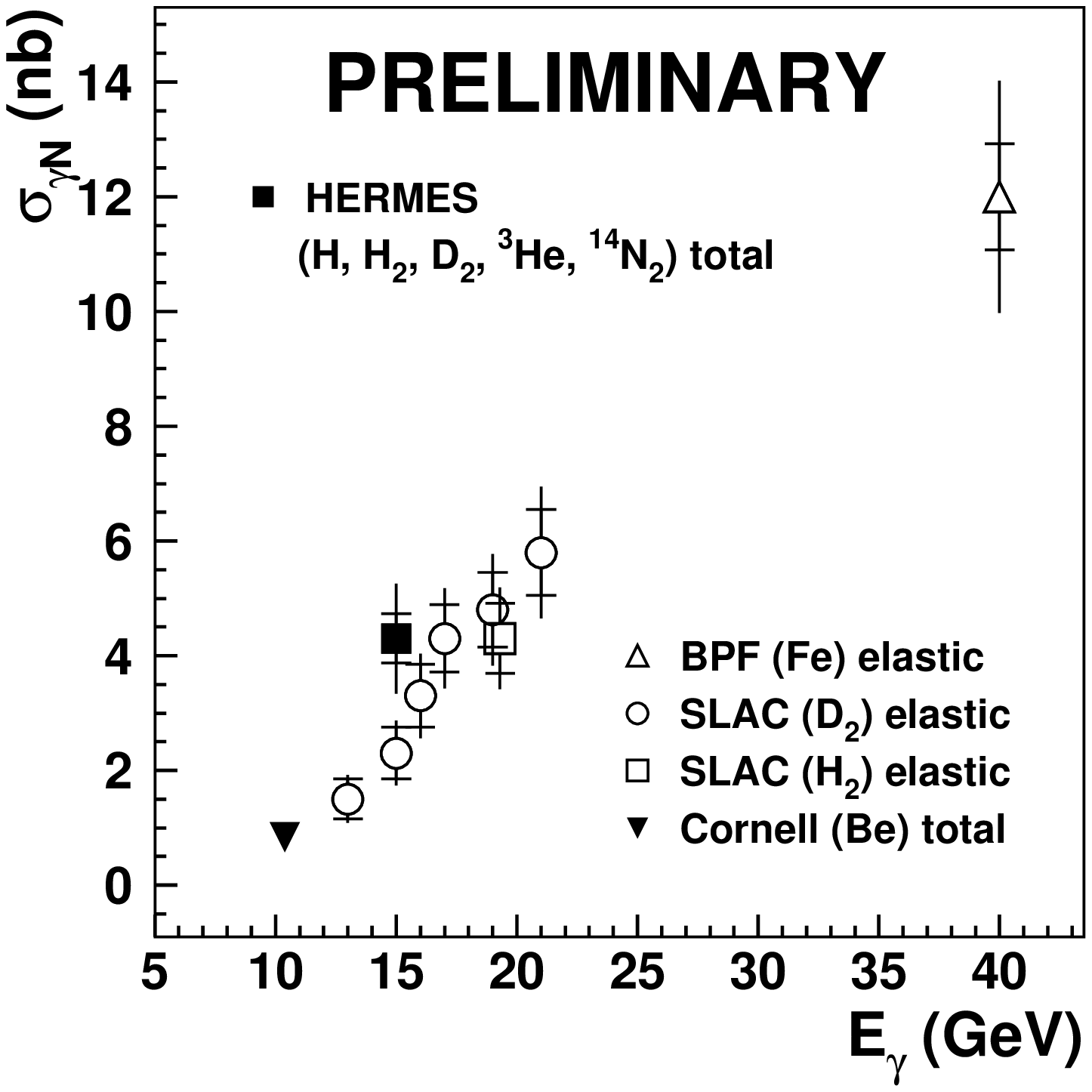,width=6cm,height=6cm} 
 \end{center}
 \caption{
  HERMES $\jp$ photoproduction cross section.
   \label{fig:hermes}
 }
\end{figure}
   The HERMES experiment at HERA presented\cite{Aschenauer:1998mu} preliminary 
   results (Fig.\ref{fig:hermes}) for open and hidden charm photoproduction at
   threshold.  They
   expect to collect data in 1999 with an upgraded detector, and to provide a 
   measurement of open charm cross section at threshold, with the ultimate
   goal of    extracting the gluon momentum distribution
   $G(x)$. Theoretical predictions of cross sections at  threshold 
   suffer from major
   difficulties (the size of $\alpha_S$, the role of higher order
   corrections, etc.). For the case
   of charmonium production at threshold, NLO predictions do
   exist\cite{Maltoni:1997pt,Petrelli:1997ge}. New theory results
   may come from the utilization of novel methods (resummation of NLO
   logarithms) developed for the high-energy region
   \cite{Mangano99,Bonciani:1998vc}.
%   (COMMENTI DI MANGANO ?????)\cite{Frixione:1997ma}
\par
   New results on $G(x)$ have come from H1 and ZEUS. H1\cite{Adloff:1998vb}
   determines the NLO gluon momentum distribution for 
   $7.5\cdot 10^{-4} < x < 4\cdot 10^{-2}$ from DIS and direct detection of
   $D^*$'s photoproduced in the final state. Results on $G(x)$ agree
   with distributions found from scaling violations of the proton structure
   function. ZEUS\cite{Deppe99} finds    the $D^*$ differential
   cross section  well described by NLO QCD calculations, with {\it
   massless} calculations\cite{Kniehl:1996we}   performing better than {\it
   massive}    calculations\cite{Frixione:1995dv,Frixione:1995qc}.
   ZEUS also remeasured\cite{Breitweg:1999ad,Bailey99}
   with higher statistics the charm contribution $F^{\ccb}_2$ to the proton
   structure function $F_2$, finding that at low-$x$ values a very large
   $(\sim 30\%)$ fraction of DIS events contains open charm states, unlike 
   EMC fixed target results at high-$x$. For a recent summary of HERA
   heavy quark results, see ref.\cite{Naroska98}.
   Finally, a further observable   that  can be computed by NLO QCD is
   the probability $g_{\ccb}$ of $\ccb$    pair production by gluon
   splitting $(\epem \rarr q\bar{q}g, g\rarr 
   Q\bar{Q})$. New OPAL preliminary results\cite{Abbiendi:1999sx}
   $g_{\ccb}=3.20\pm0.21\pm0.38)\times 10^{-2}$ are higher than theoretical
   estimates, as also recent measurements of $g_{b\bar{b}}$ by ALEPH
   and DELPHI. 
%
%     {\bf ALEPH} $D^{**}$ production fractions in B decays {\bf OPAL}
%     production  $c \bar{c}$ pairs {\bf ZEUS} $D^*$ production @DIS99 and
%     @PHOTON99 {\bf H1} $D^*$ production ??????????here ??????
%
\par
%
%     {\bf D0} $J/\psi$ suppression {\bf NA50} @HQ98 and @QM99 $J/\psi$
%     suppression {\bf WA92} @HQ98 $J/\psi$ suppression {\bf NUSEA}
%     $\sigma(J/\psi, \psi^\prime)$ {\it vs.} A, single-$\mu$ open charm
%    ?maybe in $\pspr$ suppression ????? 
  In 1992 the CDF collaboration\cite{Abe:1992ww}
 discovered that the production of  $\jp$
 and $\pspr$ in  $p\bar p$ collisions  was  enhanced by a
 factor of fifty with respect to predictions of the 
 color-singlet model, which stated that produced $\ccb$ pairs would
 dress into the observed  charmonium state by  keeping their  quantum
 numbers, i.e., by rearranging their colors without gluon emission.
 To explain the result, a color-octet model was proposed in
 which the $\ccb$ pair dressed in a charmonium hadron by emitting a soft
 gluon. The color-octect model predicts absence of polarization for
 charmonium production, since the initial polarization of $\ccb$ pairs is
 destroyed by the radiation of gluons.  Data by fixed target (WA92 at CERN), 
 and neutrino  (NUSEA) experiments on the polarization of $\jp$ do confirm 
 the color-octet prediction,   while relevant
 polarization is observed in  the HERA $q^2$ regimes\cite{Naroska98}, 
 and at the collider (CDF),  in
 agreement with the color-singlet model.
 The issue of charmonium production is dealt with in great detail in these
  proceedings\cite{Zieminski99}.
\par
 Charmonium production is also investigated on the very distant field of
 relativistic heavy ion collisions, 
 where the NA50 experiment\cite{Abreu:1999qw}
 using  1996 data (158~GeV/nucleon Pb beams on Pb target) provided
 circumstancial evidence for charmonium suppression, which may be explained
 by the onset of a quark-gluon plasma regime. 
\begin{figure}
 \begin{center}
  \epsfig{file=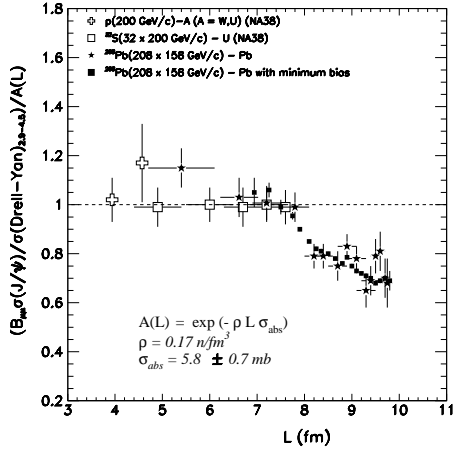,width=6cm,height=6cm} 
 \end{center}
 \caption{
  NA50 $\jp$ suppression in relativistic heavy ion collisions.
  \label{fig:na50}
 }
\end{figure}
They  measure $\jp$
 production relative to Drell-Yan pair production. After accounting for
 conventional  nuclear absorption, their data (Fig.\ref{fig:na50}) show
 evidence for a  suddenly lower production, due to the attracting
 force  between the $\ccb$ quarks being screened by gluons, and fewer $\ccb$
 pairs  hadronizing into $\jp$.
\par
Those uncorrelated pieces of information taken together confirm the
important role of gluons in the context of charmonium production dynamics. 
% (PROBLEMA DEL NUMERO DI C PRODUCED IN B DECAYS, MENTIONED BY GOLOWICH,
% REVIEWS BY BROWDER PEDRINI????)
\section{Lifetimes}
\begin{figure}[t]
 \vspace{5cm}
 \includegraphics{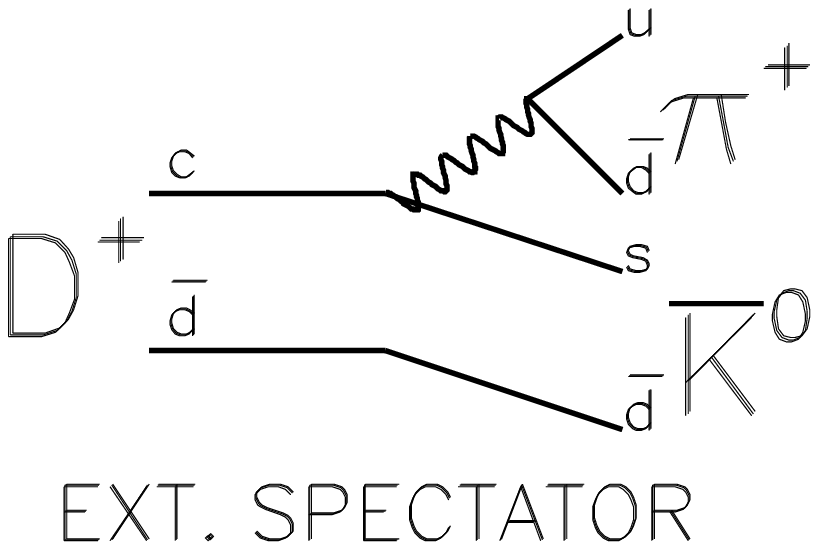}
 \includegraphics{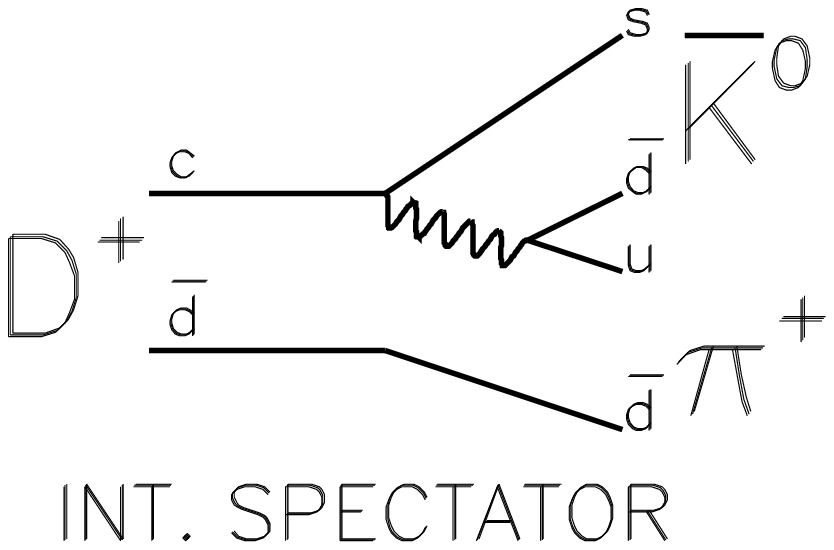}
 \includegraphics{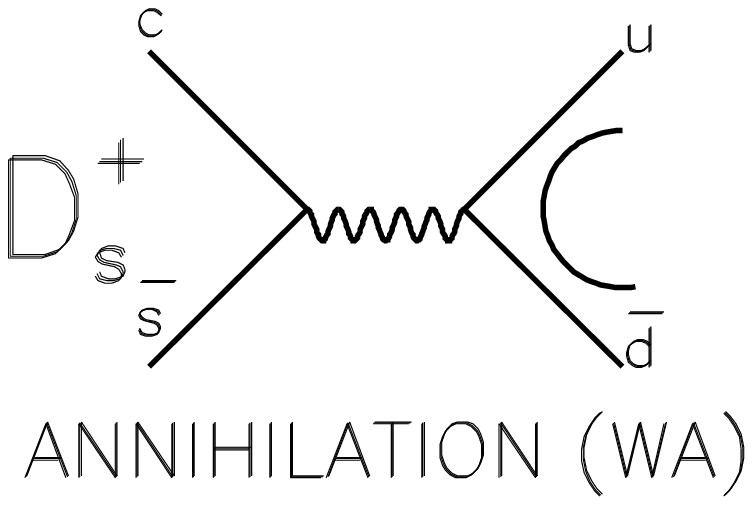}
 \includegraphics{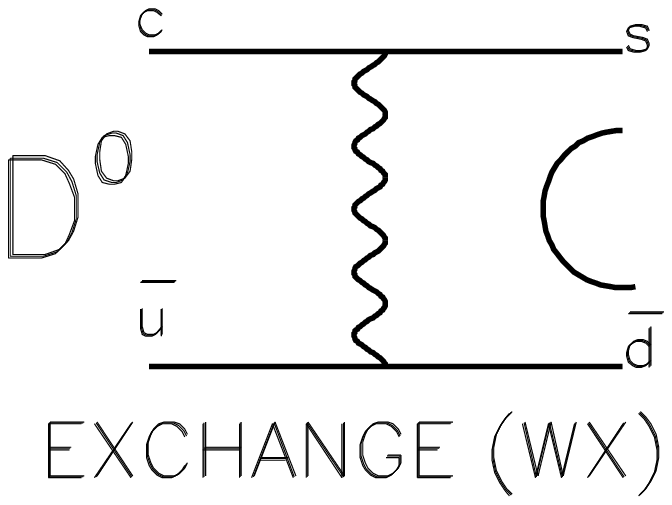}
 \caption{
 The two internal and external spectator decays are exclusive of $D^+$
  mesons and, because of the destructive Pauli interference due to the
  identical $\bar d$ quarks in the final state, they enhance the
  lifetime. The contribution to the total widths of Cabibbo-Favored (CF)
  W-annihilation and   W-exchange diagrams is, among mesons, unique to
  $D_s^+$ and $D^0$, respectively.
   \label{fig:dkgraphs}
 }
\end{figure}

\begin{figure}[p]
% \vspace{20cm}
% \special{psfile=pic99lifeproc.eps vscale=85 hscale=70 voffset=-20 hoffset=-35
%  angle=0}
 \begin{center}
  \epsfig{file=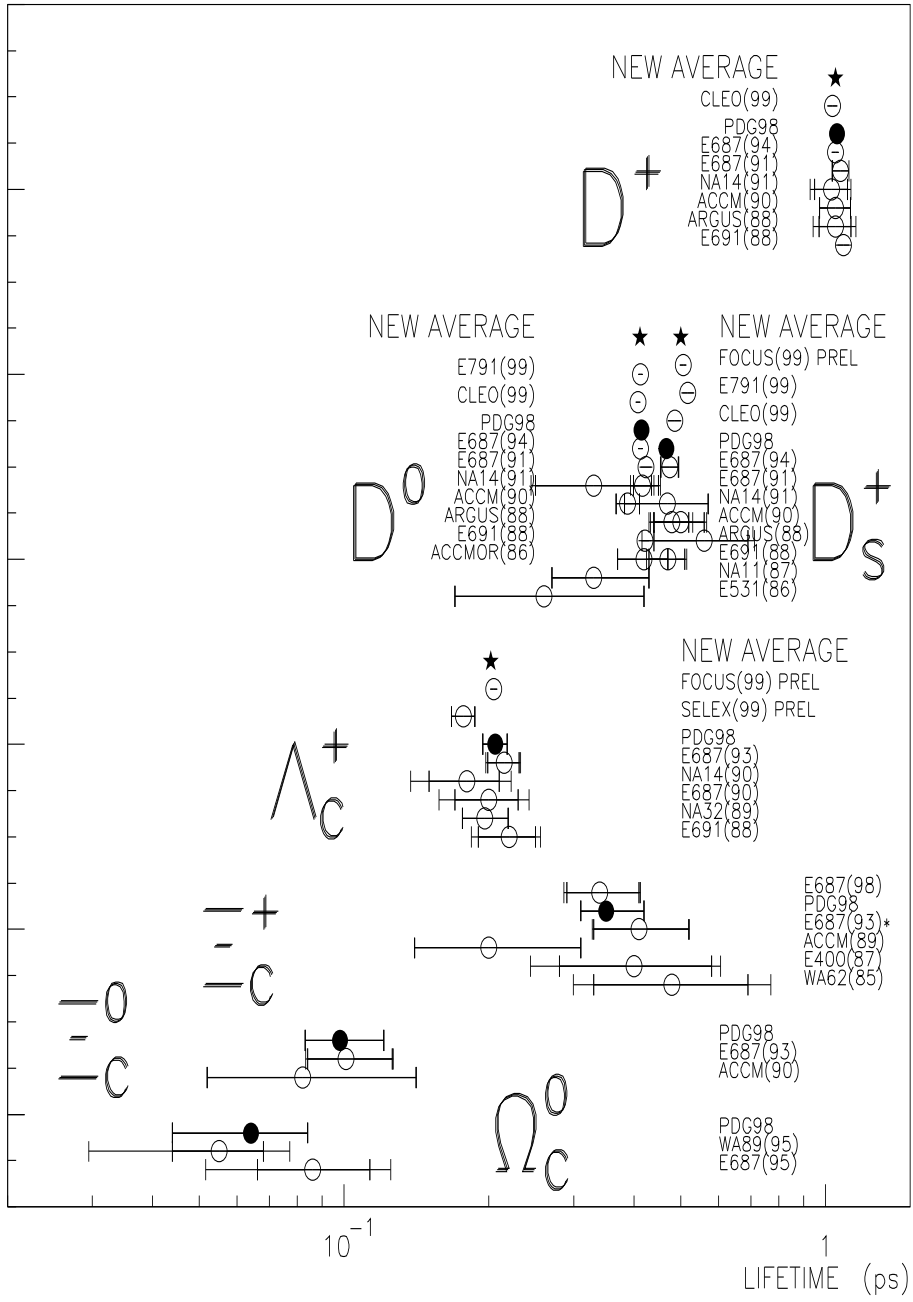,width=15cm,height=20cm} 
  \caption{Compilation of charm meson and baryon lifetimes.
    \label{fig:life}}
 \end{center}
\end{figure}

 If there were no other diagram but the spectator and no QCD effects 
 causing charm hadrons to decay, we would have one lifetime for all
 states. The wide range of lifetimes measured (Fig.\ref{fig:life}) shows
 the extent to which this is not the case.
 The total width is written as a sum of the three possible classes of decays
 $$ \tau \equiv \frac{\hbar}{\Gamma_{Total}} \equiv
 \frac{\hbar}{\Gamma_{Semilept}+\Gamma_{Nonlept}+\Gamma_{Lept}}$$ 
 The partial width $\Gamma_{Semilept}$ is universal (equal) for  $D^0$ vs
 $D^+$  (an assumption  experimentally verified within 10\%) as a
 consequence of isospin invariance, and for $D^+_s$ vs $D^0$ on the basis of
 theoretical arguments.
 The partial width $\Gamma_{Lept}$ is small due to the helicity suppression.
%resulting lepton and antineutrino must  both be either RH or LH in order to
%conserve angular momentum. However the V-A nature of weak interaction
%requires LH particles and RH antiparticles.
 Therefore, all differences
 experimentally found should be caused by $\Gamma_{Nonlept}$. 
 Lifetimes are a window on decay dynamics: conventional explanations  of
 differences among charm hadrons lie in the interplay among the spectator,
 W-exchange, and  W-annihilation diagrams (Fig.\ref{fig:dkgraphs}).
 The large difference in lifetimes for $D^0$ and $D^+$ is conclusively
 explained as being due to the presence of external and internal spectator
 diagrams. Instead, the $D^+_S$ lifetime as it appears in
 PDG98\cite{Caso:1998tx}
 is only 
 different from $D^0$ at the $3\sigma$ level, i.e., $\tau_{D^+_s}/\tau_{D^0}
 = 1.12  \pm 0.04$.
\par
 The PDG98 measurements of charm lifetimes are dominated by old
 fixed-target photoproduction E687 experiment  results. Besides new results
 from fixed-target experiments, a new player in the lifetime game in
 1998-1999 was 
 the $\epem$ experiment \c2. Their lifetime measurements\cite{CLEO99a}
 (relative to $3.7\,{\rm fb}^{-1}$, i.e., about  40\% of their present
 data set) were made 
 possible by the  implementation of a double-sided Si vertex
 detector, which also has the beneficial effect of improving
 $D^*$-tagging by a better definition of the soft pion track.
% MAIN SOURCES OF SYSTEMATICS????
%\begin{figure}[t]
% \epsfig{file=CLEO99aps.20.eps,width=7cm,height=9cm} 
% \caption{ ref: C.~Sedlack APS March 99
%   \label{fig:recon}
% }
%\end{figure}
 New results\cite{ch98} are shown in Tab.\ref{tab:life},  with new world
 averages. Although \c2 precision  is at the level of E687, their continuous
 building up of statistics, a possible better understanding of the
 systematics of their new detector, and the planned CLEO~III implementation
 of RICH particle ID may help them become competitive with
 fixed-target experiments in the future.
\par
 The most relevant  new information comes from the E791\cite{E79199b} and
 FOCUS\cite{FOCUShf8}
 measurements of $D_s$ lifetime, which reduce the error on the ratio
 with the $D^0$ lifetime
\be
     R_\tau \equiv \tau_{D^+_s}/\tau_{D^0} = 1.22 \pm 0.02 
\ee
 which is now ten standard deviations away from unity, indicating that
 although not dominant, the WA diagram is significant. In an
 approach based on Wilson's OPE\cite{Wilson:1969ey} (where the interaction
 is factorized into three 
 parts -- weak interaction between quarks, perturbative QCD corrections,
 non-perturbative QCD effects),
 the decay rate is expanded in the heavy quark masses\cite{bi96}
\bea
%\Gamma (H_Q\rarr f) & = & \frac{G_F^2 m_Q^5}{192\, \pi^3} |KM|^2  
%  \Bigl[ c_3^f \langle H_Q|\bar{Q} Q| H_Q \rangle +
%   c_5^f \frac{\langle H_Q|\bar{Q} i \sigma \cdot GQ| H_Q \rangle}{m^2_Q}+
%  \nonumber \\[4pt] 
%&& + \sum_i c^f_{6,i} \frac{\langle H_Q|(\bar{Q} \Gamma_i q) (\bar{q}
%   \Gamma_i Q)| H_Q \rangle}{m^3_Q} + 
%   {\cal O} (1/m_Q^4) \Bigr]
  \Gamma (H_Q\rarr f) & = & \frac{G_F^2 m_Q^5}{192\, \pi^3} |KM|^2  
 \Bigl[ A_0 + \frac{A_2}{m^2_Q} + \frac{A_3}{m^3_Q} + {\cal O}(1/m_Q^4) \Bigr]
   \label{eq:bi96}
\eea
 Each term has a simple physical meaning: the leading operator 
%$\bar{Q}Q$ 
 $A_0$  contains the spectator diagram contribution; 
%$\langle H_Q|\bar{Q} i \sigma \cdot GQ| H_Q \rangle$  
 $A_2$ is the spin interaction  of the heavy quark with
 light quark degrees of freedom inside the hadron; $A_3$, the PI, WA, WX
 contributions. 
 A  description of OPE goes beyond the scope of this review; interested
 readers are addressed to excellent review\cite{be97}.
 The OPE model  predicts $R_\tau=1.00-1.07$ if the WA operator
 does not contribute. If it does, the maximum effect predicted
 is $\pm 20\%$, i.e., $R_\tau=(0.8-1.27)$. The world average found is
 presently quite 
 at the limits of the OPE predictions, and it could be used as a
 constraint to better define the WA operator, which also intervenes in
 semileptonic beauty decays\cite{Bigi99}.
\par
% cumalat at sif: ratios lifetimes=ratios diagrams, using also precise DCSD
% measurement,
% can wqe estimate gamma(WX)/gamma(spect) and gamma(WA)/gamma(spect) ?
%\par
%FULL DISCUSSION BE97 P.95 INTERESSANTISSIMA dS/d0
%COMMENT INTERESSANTE d+/d0 P.92 SI PUO' ELABORARE
%\par 
In the baryon sector, 
the SELEX measurement\cite{Kushnirenko:1998ms} of the $\Lambda_C$ lifetime
disagrees with the PDG98 
world average dominated by E687, which is instead preliminarily confirmed by
FOCUS\cite{FOCUShf8} new high-statistics measurement.
%
%\footnote{The   FOCUS result was presented after the Ann
%  Arbor conference   and  is included here for completeness.}.
%
Finally, a  more precise measurement of $\Omega_c$ and $\Xi_c^{0}$
lifetimes is badly needed in order to confirm the lifetime pattern 
$\tau(\Omega_c^0) < \tau(\Xi_c^0) < \tau(\Lambda_c^+)< \tau(\Xi_c^+)$.
%\be
% \tau(\Omega_c^0) < \tau(\Xi_c^0) < \tau(\Lambda_c^+)< \tau(\Xi_c^+) 
% \label{eq:be97}
%\ee
% LIFETIMES RESULTS SINCE PIC~98 
%{\bf E687 } PL {\bf B427} (1998) 211
%{\bf E791 } FERMILAB-Pub-99/036-E 
%{\bf E791 } PL {\bf B445} (1999) 449 
%{\bf SELEX} PRELIMINARY (@DOE99)  
%{\bf FOCUS} PRELIMINARY 50\% Ds
%{\bf FOCUS} lambdacci' preliminary HF8
%
\begin{table}[t]
 \caption{Summary of new results in charm hadron lifetimes. 
  \label{tab:life}
 }
 \footnotesize
 \begin{center}
 \begin{tabular}{|c|l|c|c|c|c|} \hline
 & Experiment & Lifetime (ps) & Events & Year & Technique \\
 \hline
 \hline
{\bf  $D^+$} &
 New Average & $1.050\pm 0.015$                    &      &    &           \\
             &
 CLEO        & $1.0336\pm 0.0221 ^{+0.0099}_{-0.0127}$ & 3777 & 98 & $e^+e^-$  \\
             &
 PDG98       & $1.057\pm 0.015$                    &      &    &  \\
 \hline
{\bf  $D^0$} &
 New Average & $0.412\pm 0.003$                    &      &    &  \\
             &
 E791        & $0.413\pm 0.003 \pm 0.004$          & 35k  & 99 & Hadroprod \\
             &
 CLEO        & $0.4085\pm 0.0041 ^{+0.0035}_{-0.0034}$          & 19k  & 98 & $e^+e^-$  \\
             &
 PDG98       & $0.415\pm 0.004$                    &      &    &  \\
\hline
{\bf  $D^+_s$} &
 New Average & $0.500\pm 0.007$                    &      &    &  \\
% next line corrected 29.6.99 HWKC wants to keep syst error out = HF8
% FOCUS prel  & $0.506\pm 0.008 \pm 0.012$          & 5668 & 99 & Photoprod \\
             &
 FOCUS prel  & $0.506\pm 0.008(stat)$              & 5668 & 99 & Photoprod \\
             &
 E791        & $0.518\pm 0.014 \pm 0.007$          & 1662 & 99 & Hadroprod \\
             &
 CLEO        & $0.4863\pm 0.0150 ^{+0.0049}_{-0.0051}$          & 2167 & 98 & $e^+e^-$  \\
             &
 PDG98       & $0.467\pm 0.017$                    &      &    &  \\
\hline
{\bf  $\Lambda_c^+$} &
 New Average  & $0.2019 \pm 0.0031$                    &      &    &  \\
              &
 FOCUS prel   & $0.2045\pm 0.0034(stat)$           & 8520  & 99& Photoprod\\
              &
 SELEX prel   & $0.177\pm 0.010(stat)$             & 1790  & 99& Hyperons\\
              &
 PDG98       & $0.206\pm 0.012$                    &      &    &       \\
 \hline
 \end{tabular}
  \vfill
 \end{center}
\end{table}
%
%QUOTE EICHTEN QUIGG AND HILL MA PERCHE' ???
%

\section{Nonleptonic weak decays}
The highest impact new measurement  is the \c2 determination of
the $\Lambda_c^+\rarr p K^-\pi^+$ absolute branching fraction. This number,
used to normalize all charm baryon branching ratios, consists of
the PDG98 average of $(5.0\pm1.3)\%$ as an average of two
model-dependent measurements, in mutual disagreement at the level of
$2-3\,\sigma$. \c2 tags charm events with the semielectronic decay of a
$D^*$-tagged $\bar{D}$, and the $\Lambda^+_c$ production with a
$\bar{p}$. Their final value is $B(\Lambda_c^+\rarr p
K^-\pi^+)=(5.0 \pm 0.5 \pm 1.5)\%$. A full discussion of the analysis
technique is reported in these proceedings\cite{bs99}.
\par
\begin{figure}[p]
 \vspace{7cm}
 \includegraphics{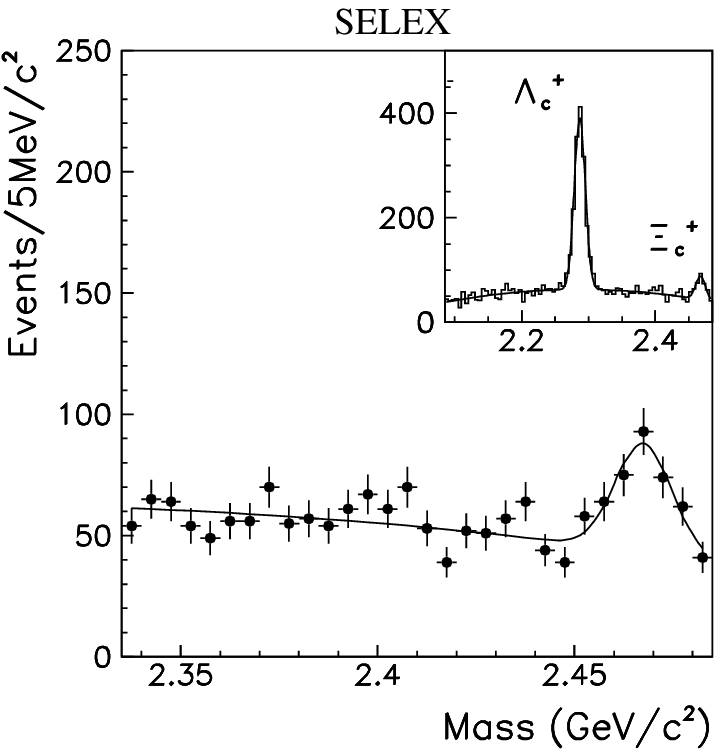}
 \includegraphics{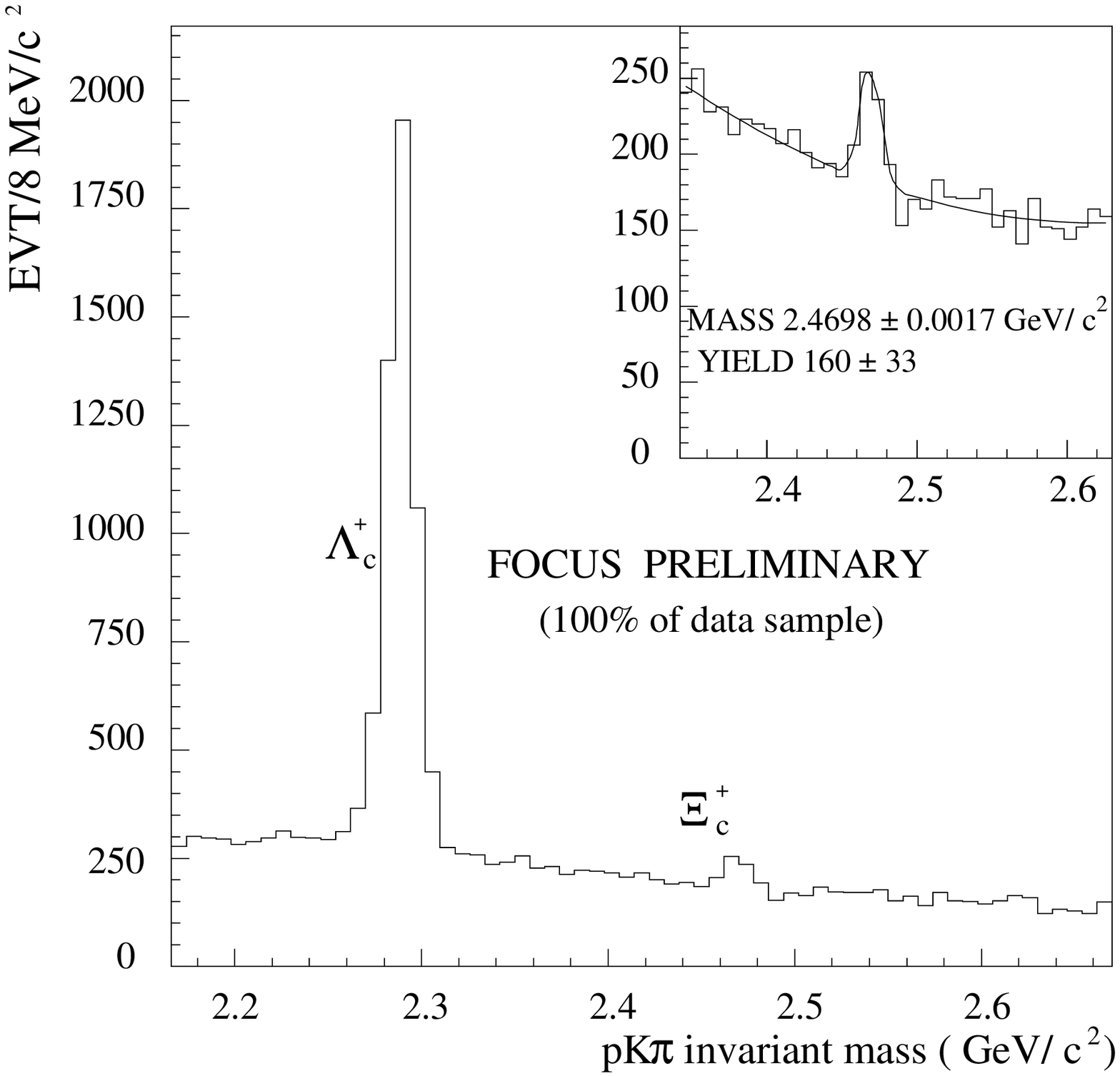}
 \includegraphics{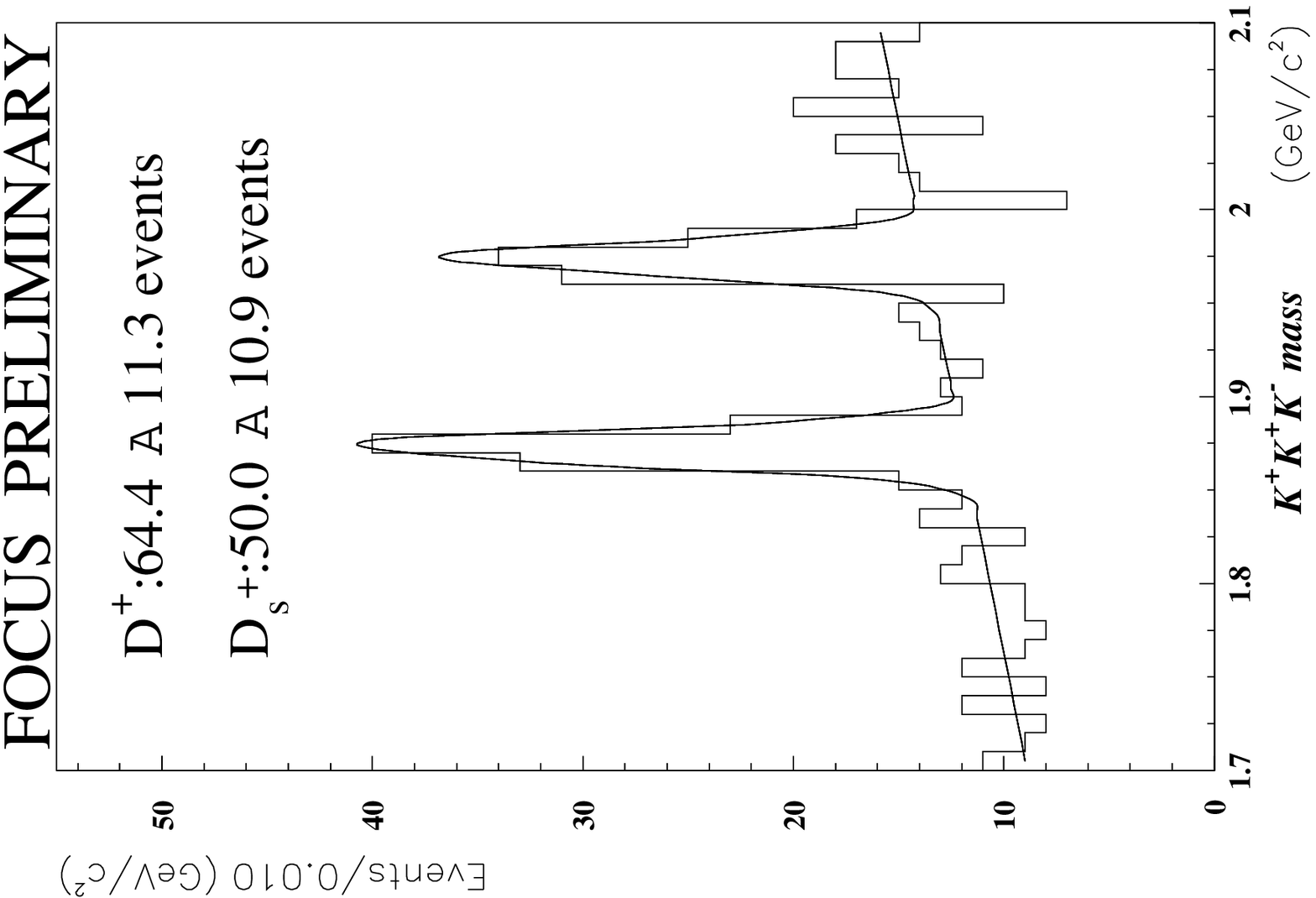}
 \caption{Observation of CS decay $\Xi^+_c
  \rarr p K^-   \pi^+$  (left and center), and  DCS decay $D^+\rarr
  K^+K^-K^+$ (right). 
    \label{fig:casc} }
\end{figure}

  First observation\cite{Jun:1999gn} (confirmed shortly
  thereafter\cite{Riccardi99}) of Cabibbo-Suppressed (CS) $\Xi^+_c
  \rarr p K^-   \pi^+$ decay by the  fixed-target, hyperon-beam  experiment
  SELEX (Fig.\ref{fig:casc}) provided 
  information on the interplay of the external W-spectator decay and
  final-state interactions (FSI). 
  These are interactions which occur in a space-time region where the final
  state particles have already been formed by the combined action of weak
  and strong forces, but are still strongly interacting while recoiling
  from each other. In charm meson decays, FSI are particularly problematic
  because of the presence of numerous resonances in the mass region
  interested\cite{br96}.
  The CS branching ratio, measured by SELEX relative to
  four--body CF decay
  $$B(\Xi_c^+\rarr pK^-\pi^+)/B(\Xi_c^+   \rarr \Sigma^+(pn)\, K^- \pi^+)=
  0.22\pm0.06\pm0.03 $$ 
  is (once  corrected for phase
  space)    compatible with the branching ratio for the only other CS decay
  well   measured, $\Lambda^+_c  \rarr p K^-  K^+$, relative to three--body CF
  decay 
  $\Lambda^+_c  \rarr p K^-  \pi^+$. This is different from 
  the charm  meson case, where branching ratios depend heavily on the
  multiplicity of   the final state, and  is interpreted as 
  confirmation of the fact that for charmed baryons, contrary to mesons, FSI
  do not play a relevant role.
 Finally, first evidence of DCS  decay $D^+\rarr K^+K^-K^+$ was
 reported\cite{Moroni99}  by
 FOCUS (Fig.\ref{fig:casc}c), which measures 
\be 
  \Gamma(D^+\rarr K^+K^-K^+)/\Gamma(D^+\rarr  K^+\pi^-\pi^+)=(1.41\pm 0.27)
  \times  10^{-4}
\ee
 Such a decay cannot proceed via a spectator diagram, since the $\bar d$
  initial state quark disappears in the final state. Possible mechanisms are
  pure WA, or Long-Distance (LD) processes including a light meson which
  strongly couples 
  to KK. In either case, a Dalitz analysis would be of extreme interest to
  possibly investigate the decay resonant structure.
 For a DCSD,  in the simplest picture one has 
 $\Gamma_{DCSD}/\Gamma_{CF} \propto \tan^4\theta_C \simeq 2  \times
 10^{-3}$. 
 Any deviation from this value  is due to effects such as interference,
  hadronization,  FSI, etc.
%
%\begin{table}[t]
% \caption{Cascade selex focus
%  \label{tab:casc}
% }
% \footnotesize
% \begin{center}
% \begin{tabular}{|c|l|c|c|c|c|} \hline
% & Experiment & BR & & & \\
% \hline
% \hline
%{\bf  $D^+$} &
% New Average & $1.050\pm 0.015$                    &      &    &           \\
% \hline
% \end{tabular}
%  \vfill
% \end{center}
%\end{table}
%
\par
 Although in principle  accessable 
 (with an important {\it caveat} being the treatment of FSI)
 by means of lattice
 methods, nonleptonic decays lack an organic theoretical
 framework rigorously descending from first principles, while
 the most interesting (two-body) decays are still largely
 undetected\cite{br96}.
 A theoretical approach that has been pointed to as
 comprehensive is  ref.\cite{bc95},\cite{bc96}, which has the merit of fully
 incorporating FSI in the prediction of two-body nonleptonic decays and
 also formulates CP-violation (CPV) asymmetries and CP-eigenstate lifetime
 differences.
%
%quote Buccella Lusignoli Pugliese FSI interaction ? \cite{bc96}
% {\bf CLEO  } Absolute $B(\Lambda_c \rarr p K \pi)$ @APS99
% {\bf SELEX } First obs of mode @HQ98
% {\bf E791 } DCSD $D^+\rarr K^+ \pi^- \pi^+$  @HQ98, $B(KK\pi \pi/K3\pi)$
% and Dalitz anal. @HQ98  
% quote bediaga comment on dalitz plot ?
% {\bf FOCUS } $\Xi_c \rarr p K \pi$ confirmation @PANIC99, DCSD  @APS99
%
%
\section{Semileptonic decays}
\begin{figure}[p]
  \vspace{5cm}
 \includegraphics{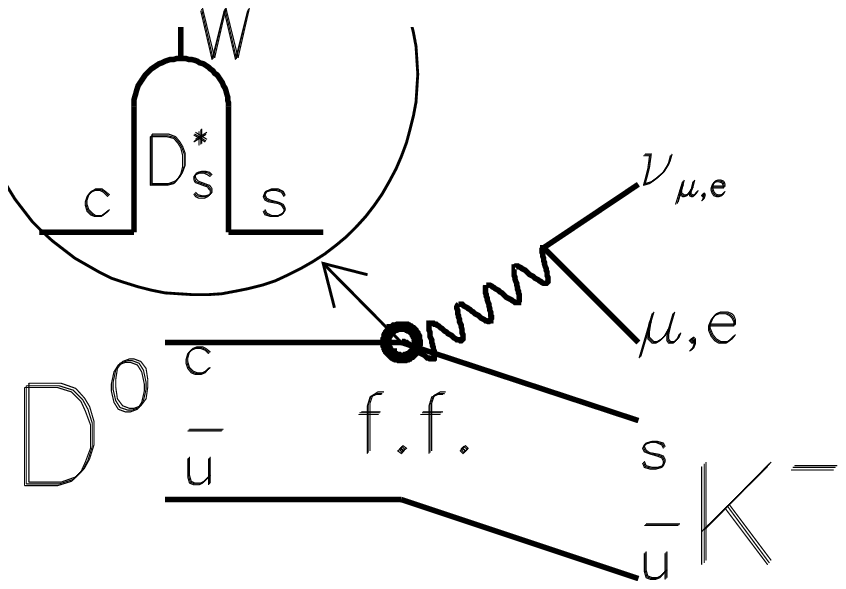} 
 \includegraphics{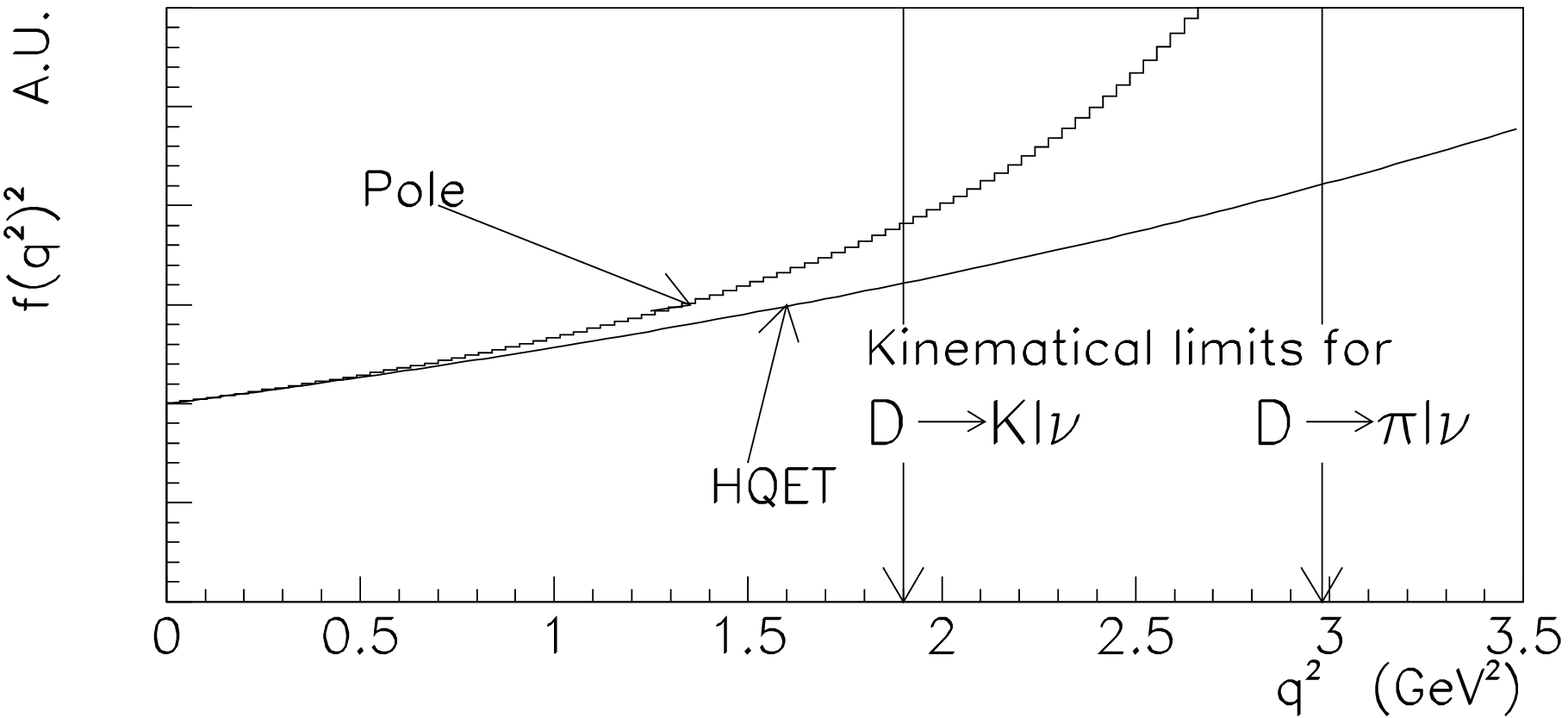} 
 \caption{Semileptonic decays of the $D^0$ meson, illustrating 
 coupling of a virtual $D_s^*(c\bar s)$ vector state which
 originates the nearest-pole dominance model (left);  $|f_+(q^2)|^2$ as a
 function of $q^2$ for pole and HQET parameterizations. The kinematic
 limits for  $K\ell\nu$ and $\pi\ell\nu$ are drawn with vertical lines
 (adapted from ref.$^8$). 
%CAVE ?????? QUOTE WISS BY HAND).
    \label{fig:slgraph} }
\end{figure}

Comprehensive older reviews of leptonic and
semileptonic decays are  in  \cite{Korner:1990qb,ri95}.
Form factors  describe dressing of $Q\bar{q}$ into a daughter hadron at the
hadronic W-vertex of the spectator decay (Fig.\ref{fig:slgraph}a). In the
simplest case of 
a charmed pseudoscalar meson decaying to a light pseudoscalar meson, lepton,
and antineutrino, the differential decay
 rate   is
\be
 \label{decrate}
 \frac{d\Gamma}{dq^2} = \frac{G^2_F |V_{cq}|^2 P^3}{24\pi^3}
 \left\{ |f_+(q^2)|^2+ |f_-(q^2)|^2{\cal O}(m_\ell^2)+... \right\}
\ee
where $P$ is the momentum of the pseudoscalar meson in the reference frame of
the charmed meson, and the $Wc\bar q$ vertex is described by only
two form factors  $f_\pm(q^2)$.
Parameterizations for $f_\pm(q^2)$ form factors inspired respectively by a
 pole dominance model and HQET\cite{Scora:1995ty} are
\begin{equation}
 f_\pm(q^2) = f_\pm(0)(1-q^2/M^2_{pole})^{-1} \qquad {\rm (Pole)}
 \label{vecmes} 
\end{equation}
\begin{equation}
 f_\pm(q^2) =  f_\pm(0)e^{\alpha q^2} \qquad {\rm (HQET)}
 \label{wisgur}
\end{equation}
The value for $M_{pole}$ is somehow arbitrarily chosen such as to be
the closest $Q\bar q$ state with the same $J^P$ as the hadronic weak
current (Fig.\ref{fig:slgraph}a).
Unfortunately (Fig.\ref{fig:slgraph}b from ref.\cite{wi98}), there is
little or no difference  between a pole or an 
exponential form in the range of small $q^2$ accessible by CF $K\ell \nu$
decays, while maximal sensitivity is allowed for CS decay $\pi\ell
\nu$. FOCUS should be able to finally measure the $q^2$ dependance by
making use of the collected sample of 5,000 $\pi\ell\nu$ semileptonic
decays. 
\par
\begin{figure}
 \begin{center}
 \epsfig{file=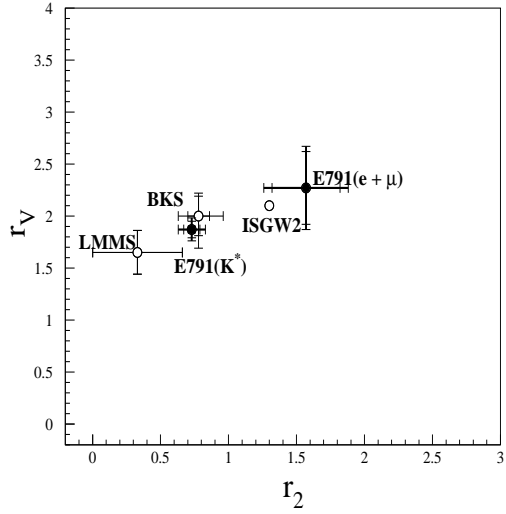,width=7cm,height=7cm} 
 \end{center}
% \vspace{5cm}
% \special{psfile=E791ff.eps vscale=50 hscale=50 voffset=-100 hoffset=-30
%  angle=0}
 \caption{E791 semileptonic vector form-factor ratios for $D^+$ and
   $D_s^+$, and comparison with predictions. 
    \label{fig:semilep}}
\end{figure}

%(COMMENTI BUTLER MERIDA WISS VARENNA BELLINI97 p.139 SEMILEP BEAUTY)
In the case of a pseudoscalar-to-vector decay there are four form factors
$(V,A_{1,3})$, $q^2$-dependent. After assuming a nearest-pole dominance
model, they are  customarily expressed via the ratios
$r_V\equiv  V(0)/A_1(0)$, $r_2\equiv  A_2(0)/A_1(0)$, with $A_3(0)$
 becoming negligible in the (questionable) limit of zero lepton mass. 
 E791 has presented new
 measurements\cite{Aitala:1998xu} of form-factor ratios for $D^+_s\rarr
\phi\ell^+\nu_\ell$, with $(\ell = e,\mu)$, which investigate the extent to
which the SU(3) flavor symmetry is valid by comparing form factors  with
 what was measured in  $D^+\rarr K^{*0}\ell^+\nu_\ell$ previously, where a
spectator $\bar{d}$ quark is replaced by a spectator $\bar{s}$ quark.
Measurements are  based on a sample of 144 electron decays and 127 muon
decays: $r_V$ is consistent with the expected 
SU(3) flavor symmetry between $D_s$ and $D^+$ semileptonic decays, while
$r_2$ appears inconsistent (Fig.\ref{fig:semilep}).
%
% IMPORTANTE COMMENTO DI JIM DA ELABORARE NON NECESSARIAMENTE PER PIC
% fare plot con ff ratios K* vs phi
%9. To put this discussion in context , you might want to refer to
%earlier measurements and controversy on the comparison of phi l nu to
%K*lnu..The e687 had about 1/2 of the statistics as this measurement by
%E791 but reached the opposite conclusion -- ie the form factors were
%consistent. CLEO also had data on this and E653 had data indicating a
%discrepancy based on only small statistics but allegedly very well
%measured. The CLEO and E653 discrepancies were on the opposite form
%factors as memory serves.
%
\par
 %OPAL \cite{Gagnon:1999nu} preprint PR259 $B(D\rarr X \ell \nu)$ 
 %\cite{Abbiendi:1998ub}
 In the $\epem$ sector, OPAL has recently produced the first
 measurement\cite{Gagnon:1999nu,Abbiendi:1998ub} of 
 the semileptonic branching ratio of charm hadrons produced in $Z^0 \rarr
 \ccb$ decays, 
 finding  $B(c\rarr \ell) = 0.095\pm0.006^{+0.007}_{-0.006}$, in good
 agreement with the ARGUS lower energy data.
%
%     {\bf E791} Form-factors ratios for $D^+\rarr K^{*0}\ell^+\nu_\ell$ and
%       $D^+_s\rarr \phi\ell^+\nu_\ell$
%     {\bf OPAL} preprint PR259 $B(D\rarr X \ell \nu)$
%     {\bf ZEUS} semilep charm decays @DIS99
%     {\bf FOCUS} promises @APS99
%
%
\section{Decay constants $f_D$ and $f_{D_s}$ in charm leptonic decays}
No experimental results have emerged on leptonic decays. Theoretical
activity on the main reasons for these studies (i.e., the
pseudoscalar decay constants $f_D$ and $f_{D_s}$) is very
intense\cite{Kronfeld98}. Lattice calculations have now converged to 
$f_{D_S} \sim 220\pm15 \,{\rm MeV}$, to be compared with the world average 
$254\pm 31\, {\rm MeV}$. On the contrary, the lattice result $f_D \sim 195
\pm 15\, {\rm MeV}$  
can only be compared with the 1988 MARK~III limit of $f_D < 290\,{\rm MeV}$.
An experiment able to study the challenging decay $D\rarr \ell \nu_\ell$ is
badly needed. An alternative model-dependent technique\cite{Rosner93} relates
the $D^{*+}-D^{*0}$ mass isosplittings to $f_D$, via the wavefunction at
the origin $|\psi(0)|^2$. The value inferred from the best isosplit measurement
\cite{Bortoletto92} is $f_D = (290 \pm 15)\, {\rm MeV} $, very distant from the
lattice computation. 
%commenti di quigg e nuovi risultati lattice p.488
%
\section{Rare and forbidden decays, CP violation.}
%
%{\bf E791 } Limits on forbidden decays (preprint  9 June 1999)
%\cite{Aitala:1999db} 
%
%{\bf FOCUS } CPV asymmetries  @KAON99
%
% quota golosh paper BLP \cite{bc95} CPV asymmetries
%
     In the charm sector, Flavor-Changing Neutral Current processes such as
     $D^+\rarr h^+ \mu^+ \mu^-$, $D^0\rarr \mu^+ \mu^-$, etc, are
     suppressed in the SM via the GIM mechanism, with predictions spanning
     an enormous range $10^{-9}-10^{-19}$. Lepton Family Number Violating
     $D^+\rarr h^+ \ell^+_1 \ell^-_2$, and Lepton Number Violating
     $D^+\rarr h^- \ell^+_1 \ell^+_{1,2}$ processes are instead strictly
     forbidden.  This is why  charm rare decays can provide unique
     information. E791 
     has presented \cite{Aitala:1999db} a set of new limits that improve the
     PDG98 numbers by a factor of 10, reaching approximately the $10^{-5}$
     region. 
%\begin{figure}[t]
% \epsfig{file=boxab.eps,width=5cm,height=10cm} 
% \caption{E791 new limits on rare D decays.
%   \label{fig:e791rare}        } 
%\end{figure}
\par
%\begin{wraptable}{r}{3cm}
% \caption{CPV   \label{tab:cpv}
% }
% \footnotesize
% \begin{center}
%\begin{tabular}{|l|r|r|} \hline
% Decay mode   & E831 prelim.  & Best limit (E791)  \\
%\hline
% $D^+ \rarr K^-K^+\pi^+$  & $-0.04\pm0.017$        & $-0.014\pm0.029$ \\
% $D^0 \rarr K^-K^+$       & $0.003\pm0.039$        & $-0.010\pm0.049\pm0.012$\\
%\hline
%\end{tabular}
%  \vfill
% \end{center}
%\end{wraptable}
     CP-violation  asymmetries in D decays are expected to occur via
     the interplay of 
     weak phases stemming from penguin, Single-Cabibbo-Suppressed
     diagrams, and a (strong) FSI phase, and are predicted at the $10^{-3}$
     level\cite{bc95}. FOCUS presented preliminary results\cite{Pedrini99}
     on CPV 
     asymmetries in the two most accessible modes ($D^+ \rarr K^-K^+\pi^+$
     and $D^0 \rarr K^-K^+$),
     which improve the current limits down to the $10^{-2}$ level. CPV in
     the charm sector still has to be discovered. The availability of large
     clean samples of fully reconstructed $D^+ \rarr K^-K^+\pi^+$ decays
     entitles one to investigate CPV by comparing phases and amplitudes 
     found in  the two CP conjugate Dalitz plots\cite{Moroni99}.
\section{$\d0d0$ mixing}
Important  new results have been presented\ on $\d0d0$
 mixing (for an updated review see \cite{sheldon99}). It is useful to 
 recall the key features  of particle-antiparticle mixing \cite{Leader96}.
 Because of weak interactions, flavor $f=s,c,b$ of a generic pseudoscalar
 neutral meson $P^0$ is not conserved. Therefore it will try and decay
 with new mass eigenstates  $P^0_{1,2}$ which no longer carry definite
 flavor ${f}$: they are new  states with different mass and lifetime
 $| P^0_{1,2} \rangle  \propto  ( p|P^0 \rangle \pm
 q|\bar{P}^0\rangle) $ 
 where complex parameters $p$ and $q$ account for any CPV.
 The time evolution of $|P^0(t)\rangle$ is given by the
 Schr\"odinger equation. After a time $t$ the probability of finding the
 state $P^0$ transformed into $\bar{P}^0$ is
\bea
 |\langle \bar{P}^0|P^0(t)\rangle|^2 \propto
 \qp2 e^{-\Gamma_1 t} [1+e^{\Delta \Gamma t}+
                 2e^{\frac{\Delta\Gamma}{2}t} \cos (\Delta m t)] 
\eea
 with definitions  $\Delta m \equiv m_1-m_2$, $\Delta \Gamma = \Gamma_1
 -\Gamma_2$ and $\bar{\Gamma} \equiv (\Gamma_1+\Gamma_2)/2$. 
 The two states will oscillate with  a rate  expressed by $\Delta
 m$ and $\Delta \Gamma$, which are naturally expressed when calibrated 
 by the  average decay rate within the parameters
 $x \equiv \Delta m/\bar{\Gamma}$ and $y \equiv
 \Delta\Gamma/(2\bar{\Gamma})$. 
\par
\begin{figure}[t]
 \vspace{2.2cm}
 \includegraphics{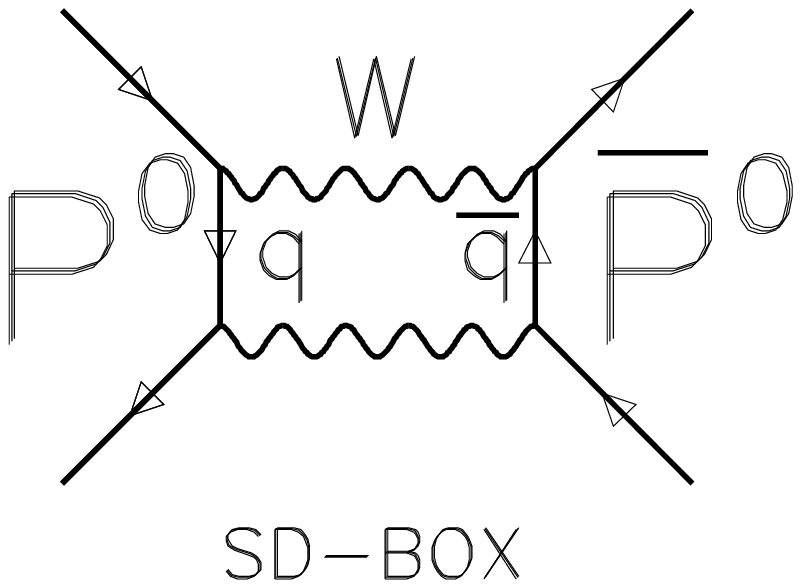}
 \includegraphics{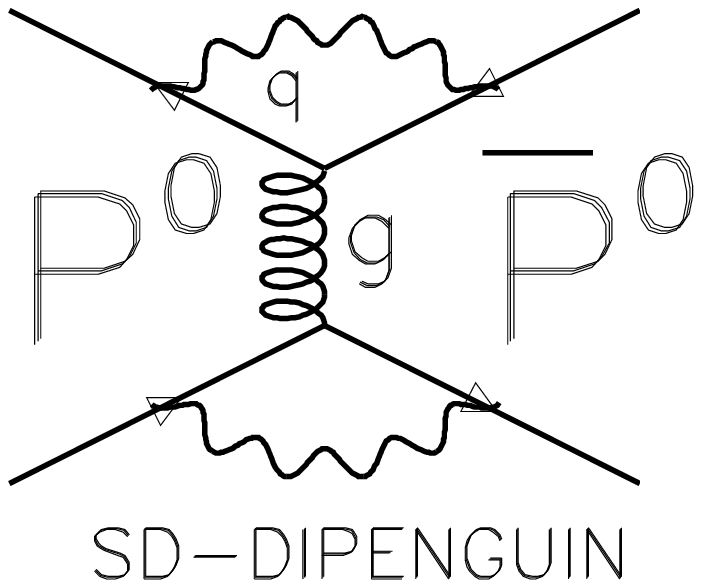}
 \includegraphics{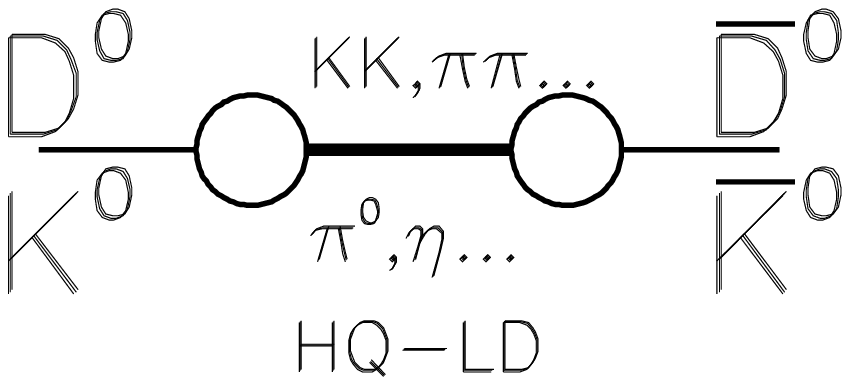}
 \caption{Box (proportional to $(m_q^2-m_d^2)^2/m_W^2m_{s,c,b}^2$),
 penguin, and 
 long-distance diagrams for mixing. \label{fig:box} }
\end{figure}
In the case of charm mesons\cite{Golowich:1995xv}, because of the
Cabibbo-favored decay mechanism and the large phase space available for
their decay, decay widths are very similar $(y\ll 1)$, 
and the time-integrated  ratio of mixed and nonmixed rates is
%\cite{Leader96} 
\be
 r\equiv \frac{\Gamma(D^0\rarr\bar{D}^0\rarr \bar{f})}{\Gamma(D^0\rarr
 f)}  =  \qp2 \frac{x^2+y^2}{2}
\ee 
\par
\begin{table}
 \caption{Box diagram contributions to mixing.
  \label{tab:mix}
 }
 \footnotesize
 \begin{center}
\begin{tabular}{|l|r|r|} \hline
 $P^0$                    & $q$  & $x\equiv \Delta m /\bar{\Gamma}$  \\
\hline
 $K^0$                    &         c              &           $0.48$  \\
 $D^0$                    & s    & $\ll 1$                     \\
% 0.481 ps-1
 $B^0_d$                    &         t              &           $0.75$   \\ 
% $B^0_s$                    &         t              &         $> 14$   \\ 
%corretto 24/11, limit is > 14.3ps-1 ==> x>22
 $B^0_s$                    &         t              &         $> 22$   \\ 
\hline
\end{tabular}
  \vfill
 \end{center}
\end{table}
 Theoretical estimates of $x$ fall into two main categories, short distance
 (SD) and heavy quark/long distance (HQ-LD): the former arise from the box
 diagram\cite{Golo4939} (Fig.\ref{fig:box}a),
 with GIM mechanism suppressing the
 charm case
 (Tab.\ref{tab:mix}) or the dipenguin diagram\cite{Petrov:1997ch},
 the latter come from QCD diagrams\cite{Golo38}
 and
 FSI\cite{Golo4939} such as
 rescattering of quarks with known intermediate light states
 (Fig.\ref{fig:box}c). 
 An important comment was made recently\cite{li97} on the possibility
 of measuring $y$ separately from $x$. Indeed, $x \neq 0$ means that mixing
 is genuinely produced by $\d0d0$ transitions (either SD or HQ-LD, or both),
 while $y \neq 0$ means that the fast-decaying component $D^0_1$ quickly
 disappears, leaving the slow-decaying component $D^0_2$ behind, which is a
 mixture of $D^0$ and $\bar{D}^0$. Infinite discussion is active on the
 extent to which the  three contributions are dominant: consensus seems to
 exist on the HQ--LD  being, in the case of charm mesons,  larger than the
 SD, and in  any case utterly small. 
 Standard Model  predictions are\cite{rpred}
\be
 x,y < 10^{-7}-10^{-3} \quad\quad r^{SM} <10^{-10}- 10^{-4} 
\ee
 still below  the PDG98 limit\cite{Aitala:1996vz} $r < 5\times 10^{-3}$.
 Any observation of $\d0d0$ mixing above the predicted level, once
  HQ--LD effects are understood, is a signal that new physics contributions
 are adding to the box diagrams\cite{Hewett:1996uc}.
Traditionally, $\d0d0$ mixing is searched for by  means of event-counting
techniques, while advances in event statistics now allow studies of
the $y$ parameter. 
\subsection{Wrong sign vs right sign counting }
 Mixing is searched for in the decay chains
\be
 D^{*+}\rarr \pi^+, \qquad D^0\rarr \bar{D}^0 \rarr
 K^+\pi^-,  K^+\pi^-\pi^+\pi^-, K^+\ell^- \bar{\nu}_\ell
\ee
 with the particle/antiparticle nature of $D^0$ at production and at decay
  given by the sign of  $\pi^+$ and $K^-$
 respectively.  
\par
 In the case of a hadronic final state, life is complicated by pollution
 of the mixing by the Doubly-Cabibbo-Suppressed Decay $D^0 \rarr K^+\pi^-$,
 proportional to $\tan^4\theta_C$.
 The  measurable $r_{WS}$ -- the rate of wrong-sign events -- has therefore
 contributions\cite{Blaylock:1995ay} 
 from DCSD, interference, and mixing
\be
 r_{WS} = \frac{\Gamma(D^0\rarr f)}{\Gamma(\bar{D}^0\rarr f)} =
 \frac{e^{-\bar \Gamma t}}{4}|\langle f|H|D^0\rangle|^2_{CF}  \qp2
 (X+Yt+Zt^2)   
\ee
\be
 X\equiv 4|\lambda|^2 \quad Y\equiv 2\Re(\lambda)\Delta\Gamma + 4 \Im(\lambda)
 \Delta m  \quad Z\equiv(\Delta m)^2 +(\Delta\Gamma)^2/4 
\ee
\be
 \lambda\equiv\frac{p}{q}
              \frac{\langle f |H|D^0\rangle _{DCS}}
                   {\langle f  |H| \bar{D}^0\rangle _{CF}} 
\ee
 The X term (pure DCS) is characterized by an exponential decay time
 behavior, unlike the Z term (pure mixing), and this feature
 can in principle be used to suppress the DCS pollution. The Y 
 (interference) term receives contributions from $\Im (\lambda)$, which can be
 nonzero if a) CPV is present, thus introducing a phase $\varphi$ in $p/q$;
 and/or b) a strong phase $\delta$ is present, due to different FSI
 in the DCS and CF decays. By assuming CP conservation, i.e.,
 $|p/q|=1$,  defining
\be
 e^{i\varphi}\equiv\frac{p}{q} \quad\quad  e^{i\delta}\sqrt{r_{DCS}}\equiv
 \frac{\langle f|H|  D^0\rangle _{DCS}}{\langle f  |H|
 \bar{D}^0\rangle_{CF}}   
\ee
 and
 measuring $t$ in units of $\bar{\Gamma}$ one can write\cite{li97} a simpler
 expression 
 for $r_{WS}$
\be
 r_{WS} \propto e^{-t} [ r_{DCS} + t^2(r/2) +
 t\sqrt{2rr_{DCS}}\,\cos\phi ]
 \label{eq:ws}
\ee
 where the interference angle is given by $\phi=\arg(ix+y)-\varphi-\delta$.
 Equation \ref{eq:ws} shows how a meaningful quote of the $r$ result must
 specify which 
 assumptions where made on the CPV and strong angles $\varphi$ and $\delta$.
 If one assumes CP invariance $(\varphi =0)$, then 
\bea
 r_{WS} \propto e^{-t} \{ r_{DCS} + (r/2)t^2 +
 (y^\prime\sqrt{r_{DCS}})t \}  \\
 y^\prime \equiv y\cos\delta-x\sin\delta \quad\quad x^\prime\equiv
 x\cos\delta+y\sin\delta 
\eea
 The alternative option in counting techniques is the use of semileptonic
 final states $K \ell \nu$, which do not suffer from DCSD pollution but are
 harder experimentally. 
\subsection{Lifetime difference measurements}
 The $y$ parameter can be determined  directly by measuring the
 lifetimes of  CP=+1 and CP=--1 final states, assuming CP conservation,
 i.e., that $D^0_1$ and 
 $D^0_2$ are indeed CP eigenstates. This would allow in principle, along with
 an independent measurement of $r$, limits to be set on $x$. The
 experimentally 
 most accessible CP-eigenstates are $K^+K^-$ and $\pi^+\pi^-$ (CP=+1),
 $K_S\phi$  (CP=--1), and $K^-\pi^+$ (mixed CP).
\begin{table}[t]  
 \caption{Synopsis of recent mixing results. CPV phase is $\varphi$, strong
 phase is $\delta$, interference angle is $\phi=\arg(ix+y)-\varphi-\delta$.
  \label{tab:mixsyn}
 }
 \footnotesize
 \begin{center}
\begin{tabular}{|l|l|l|r|r|} \hline
                      & Assumptions    & Mode      &$N_{RS}$&Result (\%)  \\
\hline
{\bf ALEPH}\cite{Barate:1998uy}   & No mix         & $K\pi$    & 1.0k   
                      &$r_{DCS}=1.84\pm .59 \pm .34$                      \\
{\it (95\%~CL)}       & $\varphi=0,\cos\phi=0$ &           & 
                      &$r<0.92$                                           \\
                      & $\varphi=0,\cos\phi=+1$ &          & 
                      &$r<0.96$                                           \\
                      & $\varphi=0,\cos\phi=-1$ &          & 
                      &$r<3.6$                                            \\
\hline
{\bf E791}\cite{Aitala:1996vz}    & $\varphi=0$       & $K\ell\nu$   & 2.5k
                      & $r=0.11^{+0.30}_{-0.27}\quad (r<0.50)$              \\
{\it (90\%~CL)}       &                &                 &
                      &                                                  \\
\hline
{\bf E791}\cite{Aitala:1998fg}    & No mix         & $K\pi$    & 5.6k
                      & $r_{DCS}=0.68^{+0.34}_{-0.33}\pm 0.07$           \\
{\it (90\%~CL)}       & No mix         & $K3\pi$   & 3.5k
                      & $r_{DCS}=0.25^{+0.36}_{-0.34}\pm 0.03$           \\
                      &$\varphi\neq 0$ in Y&           & 
                      & $r=0.39^{+0.36}_{-0.32}\pm 0.16$    \\
                      &                 &           & 
                      & $(r<0.85)$    \\
                      &                & $K\pi$    &
                      & $r_{DCS}=0.90^{+1.20}_{-1.09}\pm 0.44$           \\
                      &                & $K3\pi$    &
                      & $r_{DCS}=-0.20^{+1.17}_{-1.06}\pm 0.35$         \\
                      & None           &            & 
                      & $r(\bar{D}^0\rarr D^0)=0.18^{+0.43}_{-0.39}\pm 0.17$\\
                      & None           &            & 
                      & $r(D^0\rarr\bar{D}^0)=0.70^{+0.58}_{-0.53}\pm 0.18$\\
                      & No Y           &                 & 
                      & $r=0.21^{+0.09}_{-0.09}\pm 0.02$                 \\
\hline
{\bf E791}\cite{Aitala:1999nh}    & $\varphi=0,\delta=0$ & $KK$    & 6.7k
                      & $\Delta\Gamma=0.04\pm0.14\pm0.05\,{\rm ps}^{-1}$   \\
{\it (90\%~CL)}       &                &    $K\pi$  & 60k
                      & $(-0.20<\Delta\Gamma<0.28)\,{\rm ps}^{-1}$         \\
                      &                &              & 
                      & $y=0.8^{+2.9}_{-1.0} \quad (-4<y<6)$ \\
\hline
{\bf CLEO}\cite{Selen99}    & $\varphi=0,\delta=0$ & $\pi\pi$    & $475$
                      &                                              \\
{\it (90\%~CL)}       &                &    $KK$        & 1.3k 
                      & $y=-3.2\pm 3.4 \quad (-7.6<y<1.2)$                              \\
$5.6\,{\rm fb}^{-1}$  &                &    $K\pi$      & 19k
                      &                                                   \\
\hline
{\bf CLEO}\cite{Artuso:1999hy}    & No mix         & $K\pi$    & 16k
                      & $r_{DCS}=0.34\pm 0.07 \pm 0.06$                  \\
{\it (95\%~CL)}       & $\varphi=0,\delta\neq0$       &              & 
                      & $r_{DCS}=0.50^{+0.11}_{-0.12}\pm 0.08$           \\
$9\,{\rm fb}^{-1}$    & $\varphi=0,\delta\neq0,x^\prime=0$       &              & 
                      & $y^\prime =-2.7^{+1.5}_{-1.6}\pm 0.2$  \\
                      &           &              & 
                      & $(-5.9 < y^\prime < 0.3) $           \\
                      & $\varphi=0,\delta\neq0,y^\prime=0$       &              & 
                      & $x^\prime =0\pm 1.6 \pm 0.2$           \\
                      &                                   &              & 
                      & $|x^\prime| < 3.2 \quad (r<0.05)$           \\
\hline
\end{tabular}
  \vfill
 \end{center}
\end{table}
%
%
%
%\begin{figure}[p]
% \epsfig{file=ALEPHmix.eps,width=5cm,height=5cm} 
% \epsfig{file=E791d0life.peak.eps,width=5cm,height=5cm} 
% \epsfig{file=E791hadmix.eps,width=5cm,height=5cm} 
% \epsfig{file=E791slmix.eps,width=5cm,height=5cm} 
% \epsfig{file=smumass_rs_and_bg.eps,width=2cm,height=2cm} 
% \epsfig{file=CLEOmixkk.eps,width=3cm,height=3cm} 
% \epsfig{file=CLEOmixksphi.eps,width=3cm,height=3cm} 
% \caption{
%   blabla ALPEH CLEO E791 FOCUS
%   \label{fig:mixsyn} 
% }
%\end{figure}
%

%
\subsection{Mixing results and projections}
 The most recent mixing results in
 Tab.\ref{tab:mixsyn} are compiled from ALEPH (out
 of $4\times 10^6$ 
 hadronic $Z$  decays) and  E791 ($2\times10^5$
 reconstructed decays) recently published results, as well as
 \c2 preliminary results (lifetime difference $(5.6\,fb^{-1})$ and hadronic
 counting $(9.0\,fb^{-1})$).
 Once compared with the same assumptions, measurements are
 consistent, with the exception of a mild discrepancy in the ALEPH
 measurement of $r_{DCS}$. \c2's best limit on $y^\prime$ corresponds to
 $r<0.05\%$ if $x^\prime=0$.  Progress should come from \c2 (semileptonic
 counting and lifetime differences including the $K_S \phi$ decay mode) and 
 FOCUS, which both project sensitivities around $y^\prime <1\%$,
 corresponding to   $r<0.005\%$.
%\cite{wiss99}\cite{pedrini99}\cite{sheldon99}
% NO !  ???comment on y limit and BLP paper???
 A compilation of predictions on $\d0d0$ mixing was recently
 presented\cite{Nelson:1999fg}, and
 it was pointed out how the \c2 limit  already rules out a lot of them.
% The limit on $y$ is consistent? with the prediction based on the study of
% FSI bla bla \cite{bc96}. as obtained from an explicit sum over exclusive
% modes deltagamma/gamma sim (1.5+i 0.0014) cdot 10-3, gamma cp+1 gt Gamma
% cp-1. 
\par
 It is  clear that,  in order to further reduce the limit on $r$, one
 should await results from  B-factories. Mixing and CPV are the very topics
 which sorely  experience the lack of a concrete project of $\tau$--charm
 factory. 
\section{Spectroscopy}
 A specialized review of the many new results
 available and their implications for Heavy Quark Symmetry predictions was
 given at the  conference\cite{bs99}, so  I  shall present only an overall
 picture, 
 referring the  interested reader to details in D.~Besson's paper.
\par
 With the high-statistics, high-mass resolution experiments attaining
 maturity, focus has been shifted from the ground state ($0^-$ and $1^-$) 
 $c\bar q$ mesons and $(1/2^+$ and $3/2^+)$ $ cqq$ baryons to the orbitally-
 and, only very recently, radially-excited states\footnote{In the past,
 these excited states were called generically and 
 improperly  $D^{**}$.}. 
 An organic and consistent theoretical framework for the spectrum of
 heavy-light mesons is given by the ideas of Heavy Quark Symmetry (HQS), 
 later generalized by Heavy Quark Effective Theory in the QCD framework.
 The basic idea (mediated from the $JJ$ coupling in atomic
 physics) is that in the limit of infinite heavy quark mass: a) the much
 heavier quark does not contribute to the 
 orbital degrees of freedom, which are completely defined by the light
 quark(s) only; and b) properties are independent of heavy quark flavor. The
 extent to which the infinite heavy quark mass limit is appropriate for
 charm hadrons is the subject of infinite discussion; however, things seem
 to work   amazingly well. For recent experimental reviews see
 \cite{fa99,ra99,Paul:1999eg}.
%\footnote{ {\it From the
% viewpoint of performing an accurate prediction of the SM, the c-quark mass
% scale presents a nontrivial challenge. Methods of chiral symmetry are not
% generally applicable because $m_c$ is too large, and the use of
% heavy-quark methodology (which incorporates both heavy quark effective
% theory (HQET) and heavy-quark
% expansions based on operator product expansions (OPE)) is questionable
% because $m_c$ may not be large enough.}\cite{go96}} 
\subsection{Mesons}
%
% commenti Quigg p.491
%$D^{*'}$ Enigma {\bf (CLEO, OPAL, DELPHI)}, First Observation of  (broad)
%  L=1,  $D_1(j=1/2)$ state {\bf(CLEO@hq98)} 
%In the limit of $\infty$ mass for the heavy quark, the spin-orbit
%interaction involving $\vec{S}_Q$ can be neglected 
%\par
%$\vec{S}_Q$  and  $\vec{j}\equiv \vec{j}_q \equiv \vec{s}_q \oplus
%  \vec{L}$ 
%are separately conserved, {\it i.e.}
%
%{\Huge $$\lim_{m_Q \rightarrow \infty} \qquad = \qquad \qquad$$}
% 
% Good quantum numbers are: $J$, $j$ and $S_Q$
%\par
%For $L=1$ $\Rightarrow$ $j\equiv s_q \oplus L = 1/2, 3/2$
%\par 
%combining with $s_Q$
%$$ j=1/2 \, J=0,1  \quad {\rm and} \quad j=3/2 \, J=1,2$$
%\par
%\par
%In the $\lim_{m_Q \rarr \infty}$ $j=1/2$ decays via an S-wave broad,
%$j=3/2$ decays via a  D-wave narrow
%\par 
    Heavy Quark Symmetry provides explicit predictions on the
    spectrum of excited charmed states\cite{Godfrey:1985xj,Godfrey:1991wj}. 
\begin{figure}
%  \vspace{5cm}
% \special{psfile=dsj.eps vscale=30 hscale=40 voffset=-10 hoffset=45
%  angle=0} 
  \begin{center}
   \epsfig{file=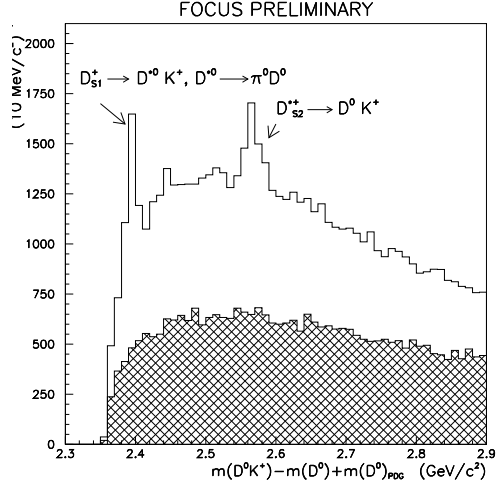,width=7cm,height=7cm,angle=0} 
  \end{center}
 \caption{The $D^0K^+$ invariant mass spectrum, showing the $D^*_{s2}$ peak
 along with the $D_{s1}(j_q=3/2)$ reflection peak due to an undetected
 $\pi^0$. Wrong-sign $(D^0K^-)$ background is shown shaded.
    \label{fig:dsj}}
\end{figure}
\begin{table}[t]
 \caption{Experimental status of (L=1, n=1) and (L=0, n=2)  $c\bar q$ and
 $c \bar s$  mesons  (adapted from ref.$^{87}$). Not  well established
%\cite{fabbri99}A MANO CITA FABBRI CAVE ???????
 states are shown in bold. Units 
 are MeV/$c^2$. Spin-parity assignment for $ D^*_{s2}$ is not pinned
 down yet, therefore this state is quoted as  $ D_{sJ}(2573)$ in PDG98.
 Theory predictions are taken from ref.$^{81,93}$.
\label{tab:dst}
 }
 \footnotesize
 \begin{center}
 \begin{tabular}{|l|c|c|c|c|c|c|} \hline
 $j_q$     & $1/2$    & $1/2$  & $3/2$ & $3/2$   & $1/2$ & $1/2$  \\ 
 $J^P$     & $0^+$    & $1^+$  & $1^+$ & $2^+$   & $0^-$ & $1^-$  \\
 $L$       & 1        &  1     &  1    &   1     &   0   &   0    \\ 
 $n$       & $1$      & $1$    & $1$   & $1$     & $2$   & $2$    \\ 
\hline
      & ${\bf D_0^*}$ & ${\bf D_1}$  & $D_1$ & $D_2^*$ & ${\bf D^\prime}$ 
&  ${\bf D^{*\prime}}$ \\  
Mass  exp. &          & $2461^0~^{+42}_{-35}$ &
$2422^0,2427^\pm$&$2459^{0,\pm}$  &  &$2637\pm6$  \\  
Mass  th.  & 2400     & 2490      & 2440  &  2500 &  2580    & 2640\\
Width exp. &          & $290^0~^{+104}_{-83}$ &$19^0,28^\pm$    &
$23^0,25^\pm$  &   & $<15$   \\
Width th.  & $>170$   & $>250$       &$20-40$  & $20-40$ &   & $40-200$ \\
Decay Mode & $D\pi$   & $D^*\pi$              & $D^*\pi$        &
$D\pi,D^*\pi$ &   & $D^*\pi\pi$  \\
\hline
           & ${\bf D_{s0}^*}$ & ${\bf D_{s1}}$ &  $D_{s1}$ & $D_{s2}^*$ & ${\bf
 D_s^\prime}$  &  ${\bf D_s^{*\prime}}$            \\  
Mass  exp. &          &        &  $2535^\pm$ &  $2573^\pm$ & & \\  
Mass  th.  & 2480     & 2570   & 2530   & 2590  & 2670   & 2730   \\
Width exp. &          &        & $<2.3^\pm$ & $15^\pm$  &  &  \\
Width th.  &          &        & $<1$   & $10-20$  &            &         \\
Decay Mode &          &        & $D^*K$      & $DK$        & & \\
 \hline
 \end{tabular}
  \vfill
 \end{center}
\end{table}
    In the limit of infinite heavy
    quark mass, the spin of the heavy quark ${\bf S_Q}$ decouples from the
    light     quark degrees of freedom (spin ${\bf s_q}$ and orbital ${\bf
    L}$), with ${\bf S_Q}$ and ${\bf j_q} \equiv {\bf s_q} +  {\bf
    L}$ the conserved quantum numbers. Predicted 
    excited states are formed 
    by combining ${\bf S_Q}$ and ${\bf j_q}$. For $L=1$ we have $j_q=1/2$ and
    $j_q=3/2$ which, combined with  $S_Q$, provide prediction for two
    $j_q=1/2$ (J=0,1) states, and two $j_q=3/2$ (J=1,2) states. These
    four states\footnote{
      Common nomenclature for excited states is
      $D_J^{*(\prime,\prime\prime,...)}$, where J is the total angular momentum
      (spin+orbital),  $(\prime,\prime\prime,...)$ indicates radial
      excitations, and   $*$ indicates  natural $(0^+,1^-,2^+,...)$ $J^P$
    assignment.} 
    are named respectively $D_0^*$, $D_1(j_q=1/2)$,
    $D_1(j_q=3/2)$ and $D_2^*$. 
%(Fig.\ref{fig:dspec})
   Finally, parity and
    angular momentum 
    conservation force the $(j_q=1/2)$ states to decay to the ground
    states via S-wave transitions (broad width), while  $(j_q=3/2)$ states
    would     decay via D-wave (narrow width).
 \par
 Such a pattern was recently beautifully borne out  by the CLEO evidence
 for  the $D_1(j_q=1/2)$ broad state\cite{Artuso:1999hy}. 
 An open question remains the
 1997 DELPHI observation\cite{ab98} of first radial excitation $D^{*\prime
 +}$ in the $D^{*+}\pi^-\pi^+$ final state, not confirmed either by OPAL
 \cite{opal98} or by CLEO\cite{Timm99}, and questioned by theory
 predictions\cite{Melikhov:1998jm}.
 The situation of
 our present  knowledge of excited $D$ mesons is shown in
 tab.\ref{tab:dst}. A major confirmation for HQS would also be the
 observation of the missing $(L=1, j_q=1/2, J^P=0^+)$ $D_0^{*}$ broad
 state. Finally, very little is known on  excited $(c\bar s)$ states
 (Fig.\ref{fig:dsj}).

 \subsection{Baryons}
 In the framework of SU(4) at ground state  we expect nine $cqq$
    $J^P=1/2^+$ baryons (all of them detected after the recent \c2 observation
    of $\Xi^\prime_c$) and six $cqq$ $J^P=3/2^+$
    baryons ($\Sigma_c^{*+}$ and $\Omega_c^{*0}$ remain undetected; they
    are expected to decay via the experimentally difficult channels
    $\Sigma_c^{*+} \rarr \pi^0 \Lambda_c^+$ and $\Omega_c^{*0} \rarr \gamma
    \Omega_c^0$ ). All 
    the $ccq$ and $ccc$ states remain undiscovered\cite{ko94}. 
    In the HQS framework, $cq_1q_2$ baryons are considered as a system made of
    a heavy quark and a light diquark. Conserved numbers are the heavy
    quark spin ${\bf S_Q}$ and the diquark angular momentum ${\bf
    j}_{q_1q_2} \equiv {\bf L} + {\bf s}_{q_1q_2}$, where ${\bf L}$ is
    the diquark orbital momentum and ${\bf s}_{q_1q_2} \equiv {\bf s}_{q_1}
    + {\bf s}_{q_2} + {\bf l}_{12}$ the diquark total
    momentum\footnote{I  adopt for excited baryon states the nomenclature  in
    \cite{Appel:1993hn}. Thus, members of 3/2 multiplets are given a $(*)$,
    the subscript is the orbital light diquark momentum ${\bf L}$, and
    $(\prime)$     indicates symmetric quark wavefunctions $c\{q_1q_2\}$
    with respect to 
    interchange of light quarks, opposed to
    antisymmetric wavefunctions $c[q_1q_2]$.}.
\par
\begin{table}[t]
 \caption{Isospin mass splittings.
  \label{tab:iso}
 }
 \footnotesize
 \begin{center}
 \begin{tabular}{|l|l|} \hline
                 & $M(\Sigma_c^{++} -\Sigma_c^{0})\, {\rm MeV}$ \\
 \hline
FOCUS prel.           & $0.28 \pm 0.31 \pm 0.15$         \\
PDG98            & $0.57 \pm 0.23$                  \\
E791             & $0.38 \pm 0.40 \pm 0.15$         \\
%E687            &                                  \\
CLEO II          & $1.1 \pm 0.4 \pm 0.1 $                                \\
CLEO             & $0.1 \pm 0.6 \pm 0.1 $                                \\
ARGUS            & $1.2 \pm 0.7 \pm 0.3 $                                \\
\hline
 \end{tabular}
  \vfill
 \end{center}
\end{table}
 During 1998,  CLEO\cite{je99} presented evidence for the two $1/2^+$
 missing states $\Xi^{0\prime }_c$ and $\Xi^{+\prime }_c$, respectively
 $c\{sd\}$ and  $c\{su\}$, through the radiative decay  $\Xi_c \gamma$. 
 Masses measured are compatible with predictions\cite{ko94,Jenkins:1997rr}.
\par
 Only  a few of the orbitally excited P-wave $(L=1)$
  baryons have been     observed. The first doublet $\Lambda_{c1}(2593)$
  $(1/2^-)$     and $\Lambda^*_{c1}(2625)$ $(3/2^-)$ was observed several
    years ago by \c2, ARGUS, and E687. 
    In 1999 \c2 
    presented evidence\cite{al99} for the charmed-strange baryon analogous
    to $\Lambda^*_{c1}(2625)$, called $\Xi^*_{c1}(2815)$, in
    its decay to $\Xi_c \pi \pi$ via an intermediate $\Xi_c^*$ state.
    The presence of intermediate $\Xi_c^*$ instead of $\Xi_c^\prime$ is an
  indication of $3/2^-$ assignment, while HQS explicitly forbids 
  a direct transition to the $\Xi_c \pi $   ground state,
  because of  angular momentum and parity conservation.
\par
 The  last topic of this section is the mass difference between isospin states
 of charmed baryons (isosplits). 
%
% IMPORTANTE COMMENTO DI BIGI DA ELABORARE
%
%Through (sic) order 1/m_Q^2 heavy flavour baryons are actually simpler
%objects than 
%heavy flavour mesons. However in order 1/m_Q^3 -- where the most significant
%differences between meson and baryon lifetimes arise -- baryons become
%more complicated. On one hand there are more relevant operators. On the other
%hand we do not have a simple prescription like factorization baryons
%(it is meaningful for mesons only).
%At present we have to rely on quark models to calculate the baryonic
%expectation values most relevant for baryon lifetimes without the benefit of
%first principle considerations. The only guideline we have to check our
%quark models is -- baryonic spectrocopy etc.
%This is explained in my Physics Report, but also in
%Uraltsev, Phys. Lett. B 376 (1996) 303. There is also a slightly earlier
%paper by Rosner on this subject where more details are given.
%
 Lately, interest in this issue was
 revamped\cite{Rosner:1998zc,Varga:1998wp} led by the consideration
 that,  while for all 
 well-measured isodoublets one increases the baryon mass by replacing a
 $u$--quark with a $d$--quark, the opposite happens in the case of the
 poorly measured  $\Sigma_c^{++}(cuu) -\Sigma_c^{0}(cdd)$ isosplit.
 Besides the $u/d$ quark mass difference, isosplits are in
 general sensitive to   em effects, and spin--spin hyperfine interactions.
 New FOCUS results\cite{Vaandering99} are consistent with E791, and mildly
 inconsistent  with CLEO~II (Tab.\ref{tab:iso}).
 Finally, first measurement of the $\Sigma_c$ width was presented at this
 conference -- full details in D.~Besson's review.
\par
 In the baryon sector, the discovery of double-charm states would be of 
 fundamental importance. A recent theoretical work\cite{Predazzi99} shows
 how a (nearly) model--independent approach based on the Feynman-Hellman
 theorem 
 is able to compute heavy-flavor hadron masses in very good agreement with
 experiments, and to make predictions for double-charm states. As an example,
 the mass of the  $\Xi_{cc}^+$ is predicted at $\sim 3.7\,{\rm GeV}/c^2$, and
 dominant decay modes are  $D^+\Sigma^+$ and $D^+\Lambda^0$. Such a discovery
 does not seem within reach of either CLEO, or present fixed-target
 experiments.   
 \subsection{Charmonium}
 Charmonium states are produced at $\epem$ storage rings and through
 $p\bar p$ annihilation. The $\epem$ annihilation proceeds, at first order,
 via a virtual photon and only $J^{PC}=1^{- -}$ states can be directly
 formed. Nonvector states such as the $\chi_J$ states can be observed only
 via two--step processes as $\epem \rarr \psi^\prime \rarr (\ccb)+\gamma$,
 or higher order processes. On the contrary, in $p\bar p$ annihilations all
 $J^{PC}$ quantum numbers are accessible.
\par
 New data come from $\epem$ BES\cite{Harris99} and $p\bar p$ 
 E835\cite{Bettoni99}. Results include measurements of masses and widths of
 $\chi_{c0}$,  $\chi_{c2}$, $\eta_c$. A $3\sigma$ disagreement remains
 between their $\eta_c$ mass determinations. Moreover, the width
 $\Gamma(\eta_c \rarr \gamma \gamma)$ measured by E835 is nearly a factor
 of two narrower than the PDG98 world average, with which, instead, recent data
 from L3, OPAL, and BES agree.
\par
 The most intriguing puzzle in charmonium physics seems to remain the case
 for the pseudoscalar radial excitation $\eta_c^\prime \,
 (2^1S_0)$. Claimed in 1982 by Crystal Barrel at $3.594\,{\rm GeV}/c^2$, it
 was not confirmed by either DELPHI or E835 extensive searches
 ($30\,{\rm pb}^{-1}$ in the range $3.666$ to $3.575\,{\rm GeV}/c^2$). The
 E835 data-taking period at the end of 1999 sees the search for the
 $\eta^\prime_c$ approved as prioritary.
% the case of charmonnium two heavy quarks of same mass is different
% analogy LS COUPLINg vedi flf comment.
% Charmonium ($(c\bar{c}$ states) spectroscopy is studied in $e^+e^-$
% collisions, and  commenti Quigg hq98 p.491 
%\par
% $\chi_{c0}$ width, mass $\eta_c$
% {\bf E835} @MORIOND99, WHS99: $\chi_{c2}(3556)$, $\eta_c(2980)$,
%      $\alpha_S$ , $\Gamma(\chi_{cJ} \rarr p\bar{p})/\Gamma_{tot}(\chi_{cJ})$
% {\bf BES}  @DPF99: $\chi_{cJ}$, $\alpha_S$, $\eta_c$ mass and width
% {\bf OPAL} @WHS99: $\chi_{c2}$ width
%\par
%quigg: e835 chicppbarBR, no etacprime
%
\section{Conclusions and outlook}
%
%The future of charm physics comprises a year 2000 full of results from
%FOCUS, \c2, E791. Sfter that, no additional machines  taucharm no fixed
%target only compass. This may be a good time to discuss while the taucharm
%seem s to have lost momentum.
 The major news this year in charm physics comes from the advances in
 lifetime measurement technology. The $D^+_s$ lifetime is now conclusively
 measured as being larger than $D^0$, and the lifetime ratio can be used
 (along with 
 the new measurements of DCSD branching ratios) to constrain the sizes of
 the 
 WA and WX  operators. New limits on $\d0d0$ mixing also come mainly from
 the novel 
 capability of measuring the lifetimes of opposite CP eigenstates.
 The other field where impressive news comes from is spectroscopy, with
 HQS predictions being spectacularly confirmed by the  observation of an
 excited  broad meson state, while the puzzles of meson radial excitations
 and the very  existance of the $\eta_c^\prime$ still elude us.
 At the opening of the third millenium, besides results from E791,
 FOCUS, SELEX, \c2, E835, and BES,  we should expect first  data on $\d0d0$
 mixing from  the BaBar and BELLE  B-factories.
\par
 The future of post-Y2K charm physics is less clear. No new
 fixed-target data-taking periods for charm studies are planned either at
 Fermilab, or at CERN for  the near future (COMPASS
%\cite{compass} 
 commissioning is scheduled to begin in 2000, with a busy physics programme
 which includes charm muonproduction, DIS spin-physics, gluon structure
 functions, and light-quark hadronic physics).
 Despite the  significant upgrade  planned, it is not clear whether CLEO~III
 will retain  competitivity with respect to B-factories. Future experiments
 at high-energy
 hadron  machines (Hera-B, BTeV, LHC-B) will need to tame huge backgrounds
 in order to contribute to charm physics. 
 Intense workshop activity\cite{Perl:1999tq} on a $\epem$ 
 $\tau$-charm factory has not translated into an actual proposal ---
 perhaps the
 operational experience coming from low-energy, high-luminosity $(\approx
 10^{33} {\rm  cm^2 s^{-1} } )$ machines such as Frascati DA$\Phi$NE is needed.
 Interesting ideas come from a proposal for a $p\bar
 p$ collider  operating at the open charm threshold\cite{GSI}. 
 On the other hand,  a few specialized efforts have been approved, in the
 form of experiments undertaking  upgrades for specific  study of open charm
 physics (HERMES, NA50). The distant future will probably see charm
 $\nu$-production from muon storage rings. 
 Next year, the Lisbon conference will be an appropriate time to verify
 whether such a scenario had evolved.
\section*{Acknowledgments}
 This review would not have been  possible without the unconditioned help of
 the spokespersons and  analysis coordinators of the experiments which
 provided 
 me with recent data: ALEPH, DELPHI, OPAL, BES, E791,  L3, SELEX, E835, ZEUS,
 H1, NA50, HERMES, NUSEA. I especially enjoyed discussions with
 D.~Besson, D.~Asner, A.~Zieminski, and M.L.~Mangano.
 I  thank all my FOCUS colleagues for
 continuous help, especially E.E.~Gottschalk, P.~Sheldon, M.~Hosack,
 D.~Menasce, M.~Merlo, K.~Stenson, H.~Cheung, A.~Zallo, H.~Mendez,
 C.~Riccardi, G.~Boca, S.P.~Ratti, L.~Moroni, E.~Vaandering.  
  I should like to thank  I.~Bigi, J.~Cumalat, D.~Pedrini, and J.~Wiss
  for fundamental comments and my colleagues  Franco
  L. Fabbri and  Shahzad Sarwar for daily discussions on charm physics.
  Finally, I should like to thank and congratulate the International
  Advisory Committee and the Local Organizing Committee for a completely
  successful   conference. 

\end{document}